\documentclass[nofootinbib, aps, prd, a4paper, 10pt, superscriptaddress, eqsecnum, showkeys]{revtex4-2}
\usepackage[a4paper, top=2cm, bottom=2cm, left=2.5cm, right=2.5cm]{geometry}

\usepackage{amsmath}
\usepackage{amsfonts}
\usepackage{amssymb}
\usepackage{amsthm}
\usepackage{bm}
\allowdisplaybreaks[4]

\usepackage{graphicx}
\usepackage{booktabs}
\usepackage{subcaption}

\usepackage{xcolor}
\usepackage[pdfusetitle]{hyperref}
\hypersetup{colorlinks = true, allcolors = blue}

\IfFileExists{orcidlink.sty}{%
  \usepackage{orcidlink}%
}{%
  \providecommand{\orcidlink}[1]{}%
}

\begin{document}

\title{Singlet Extended Mirror Standard Model World as Dark Matter and Gravitational Waves Imprints of a High Scale Mirror Phase Transitions}

\author{Asterios T. Papadopoulos\orcidlink{0009-0008-9210-9906}}
\email{asterispap05@gmail.com,apapadopb@auth.gr}
\affiliation{Department of Physics, Aristotle University of
Thessaloniki, Thessaloniki 54124, Greece}

\author{V.K. Oikonomou\orcidlink{0000-0003-0125-4160}}
\email{voikonomou@gapps.auth.gr} \affiliation{Department of
Physics, Aristotle University of Thessaloniki, Thessaloniki 54124,
Greece} \affiliation{Center for Theoretical Physics, Khazar
University, 41 Mehseti Str., Baku, AZ-1096, Azerbaijan}

\begin{abstract}
We study a mirror Standard Model world which contains a high scale
duplicate of the Standard Model, and includes a singlet mirror
scalar extension coupled to the mirror Higgs particle. Among  the
interactions of the mirror Higgs to the mirror singlet scalar, we
include dimension-six non-renormalizable operators. We examine the
electroweak phase transition of this mirror singlet extended world
and we show that the phase transition is a strong first order
phase transition, the bubble collision of which can be detectable
by LISA, the BBO and the DECIGO gravitational wave experiments. We
also provide a rough estimate of the abundance of the mirror
singlet and we show that the whole dark matter in the Universe may
be comprised by mirror particles and atoms.
\end{abstract}

\maketitle

\section{Introduction}

In the next decade the primary focus of theoretical physicists
will be on cosmic microwave background (CMB) experiments
\cite{SimonsObservatory:2019qwx,LiteBIRD:2022cnt} and on
gravitational wave experiments
\cite{Hild:2010id,Baker:2019nia,Smith:2019wny,Crowder:2005nr,Smith:2016jqs,Seto:2001qf,Kawamura:2020pcg,Bull:2018lat,LISACosmologyWorkingGroup:2022jok}.
The CMB experiments will probe the recombination regime which in
turn constraints the primordial era, which is theorized to be
modelled by the inflationary paradigm
\cite{inflation1,inflation2,inflation3,inflation4,inflation5,inflation6}.
On the other hand, the gravitational wave experiments will probe
small frequencies that are currently out of range from LIGO/Virgo
detectors. In these frequency ranges, a plethora of phenomena may
result in the generation of a stochastic gravitational wave
spectrum, among which inflation, phase transitions and so on, see
for example
\cite{Kamionkowski:2015yta,Turner:1993vb,Boyle:2005se,Zhang:2005nw,Caprini:2018mtu,Clarke:2020bil,Smith:2005mm,Giovannini:2008tm,Liu:2015psa,Vagnozzi:2020gtf,Giovannini:2023itq,Giovannini:2022eue,Giovannini:2022vha,Giovannini:2020wrx,Giovannini:2019oii,Giovannini:2019ioo,Giovannini:2014vya,Giovannini:2009kg,Kamionkowski:1993fg,Giare:2020vss,Zhao:2006mm,Lasky:2015lej,
Cai:2021uup,Odintsov:2021kup,Lin:2021vwc,Zhang:2021vak,Visinelli:2017bny,Pritchard:2004qp,Khoze:2022nyt,Casalino:2018tcd,Oikonomou:2022xoq,Casalino:2018wnc,ElBourakadi:2022anr,Sturani:2021ucg,Vagnozzi:2022qmc,Arapoglu:2022vbf,Giare:2022wxq,Oikonomou:2021kql,Gerbino:2016sgw,Breitbach:2018ddu,Pi:2019ihn,Khlopov:2023mpo,Odintsov:2022cbm,Benetti:2021uea,Vagnozzi:2020gtf,Apreda:2001us,Schabinger:2005ei,Kusenko:2006rh,McDonald:1993ex,Chala:2018ari,Davoudiasl:2004be,Baldes:2016rqn,Noble:2007kk,Zhou:2020ojf,
Weir:2017wfa,Hindmarsh:2020hop,Han:2020ekm,Child:2012qg,Fairbairn:2013uta,LISACosmologyWorkingGroup:2022jok,Caprini:2015zlo,Huber:2015znp,
Delaunay:2007wb,Chung:2012vg,Barenboim:2012nh,Senaha:2020mop,Grojean:2006bp,Katz:2014bha,Alves:2018jsw,Oikonomou:2023bah,Gouttenoire:2021jhk,Kuroyanagi:2014nba,Ellis:2020awk,Athron:2023xlk}.
The NANOGrav collaboration in 2023 \cite{NANOGrav:2023gor}, had
detected a stochastic gravitational wave signal, but it is not yet
confirmed that the stochastic signal is cosmological, and even if
so, it is highly unlikely that this signal can be explained solely
by inflationary primordial gravitational waves
\cite{Vagnozzi:2023lwo,Oikonomou:2023qfz}.

Closely related to the primordial mysteries of the Universe are
the dark sector problems, namely the dark matter (DM) and the dark
energy problems. The DM problem is a long standing problem, and
the theoretical proposal itself dates back to Zwicky's years, when
Zwicky assumed that missing matter in the Coma cluster can explain
the rotation curves of spiral galaxies. The DM search peaked
during 1980-2010, and theorists and experimentalists hoped to find
hints of DM particles in these experiments. However, no hint of DM
was found in these experiments, so the idea of the Weakly
Interacting Massive Particles (WIMP), the main candidate for DM
during 1980-2010, was abandoned or at least the majority of
physicists lost interest towards the WIMP candidates. It seems
that the DM problem is by far more difficult to solve, but like in
the black holes case, we know that DM must be out there in
particle form, we just cannot prove it experimentally yet. In the
black holes case, we knew that black holes are at the centers of
galaxies dynamically, but we did not have observational evidence
for black holes until 2019 when the first pictures of black holes
were published by the event horizon collaboration. DM is
inherently tied up to the $\Lambda$-Cold-Dark-Matter model and in
particle form, provides consistency to cosmological evolution. It
certainly has challenges at small galactic scales, for example the
cusp-core problem in dwarf galaxies. Due to these problems,
alternative theories like Modified Newtonian Dynamics (MOND)
theories, try to mimic the effect of DM, but these theories lack
of a viable formal relativistic quantification. There are recent
works that provide such a framework, for example non-local
approaches
\cite{Deffayet:2024ciu,Boran:2017rdn,Deffayet:2014lba,Deffayet:2011sk},
but these are challenged with providing consistent explanations
for the Baryon Acoustic Observations, the CMB itself, the spin
problem of spiral galaxies, and other phenomenological problems.
An interesting viable DM candidate is provided by mirror DM
\cite{Kobzarev:1966qya,Hodges:1993yb,Foot:2004pa,Berezhiani:2003wj,Silagadze:2008fa,Foot:2000tp,Chacko:2005pe,Berezhiani:2000gw,Blinnikov:2009nn,Tulin:2017ara,Mohapatra:2001sx,
Blinnikov:1982eh,Blinnikov:1983gh,Foot:2016wvj,Foot:2014osa,Foot:2014uba,
Foot:2004pq,Foot:2001ft,Foot:2004dh,Foot:1999hm,Foot:2001pv,Foot:2001ne,Foot:2000iu,Pavsic:1974rq,Foot:1993yp,Ignatiev:2000yy,Ignatiev:2003js,
Ciarcelluti:2004ik,Ciarcelluti:2004ip,Ciarcelluti:2010zz,Dvali:2009fw,Foot:2013msa,Foot:2013vna,Cui:2011wk,Foot:2015mqa,Foot:2014mia,Cline:2013zca,Ibe:2019ena,
Foot:2018qpw,Howe:2021neq,Cyr-Racine:2021oal,Armstrong:2023cis,Ritter:2024sqv,Mohapatra:1996yy,Mohapatra:2000qx,Goldman:2013qla,Berezhiani:1995am,Oikonomou:2024geq,Oikonomou:2025jmy,Oikonomou:2026ugb}.
The mirror DM is based on the idea that the Universe is filled
with a dark copy of the Standard Model (SM), which may interact
weakly with the real SM or it may not interact at all, save
gravitationally only. This mirror DM is thus an interacting form
of DM, and can explain the Universe at cluster or supercluster
scales and even at galactic scales. This sort of interacting DM
can accommodate DM models with scale dependent EoS and this sort
of scale dependent DM can fit quite well DM-dominated galaxies
\cite{Oikonomou:2026vkp,Oikonomou:2025bsi}.

If there exists a mirror world of DM, which only interacts
gravitationally, then the question is how would this world be
detected from us? The answer lies on the stochastic gravitational
wave spectrum of this world. The mirror SM which contains a copy
of the SM, can also experience phase transitions, which can
generate a gravitational wave spectrum that can potentially be
detected by our future gravitational wave experiments. However the
phase transition must be strong first order, in order the bubbles
that are generated by the phase transition can cause a strong
stochastic spectrum of gravitational waves. To this end, in this
work we assume that a mirror DM world exists complementary to our
world, at much higher scales than the SM. In addition, we assume
that the high scale mirror DM world is equipped with a singlet
scalar, which may enhance significantly any underlying mirror
electroweak phase transition and also a higher order operator is
included in the mirror SM Lagrangian. First order phase
transitions in the ordinary SM world are known to produce bubble
collisions that induce stochastic gravitational waves
\cite{Apreda:2001us,Schabinger:2005ei,Kusenko:2006rh,McDonald:1993ex,Chala:2018ari,Davoudiasl:2004be,Baldes:2016rqn,Noble:2007kk,Zhou:2020ojf,
Weir:2017wfa,Hindmarsh:2020hop,Han:2020ekm,Child:2012qg,Fairbairn:2013uta,LISACosmologyWorkingGroup:2022jok,Caprini:2015zlo,Huber:2015znp,
Delaunay:2007wb,Chung:2012vg,Barenboim:2012nh,Senaha:2020mop,Grojean:2006bp,Katz:2014bha,Alves:2018jsw,Athron:2023xlk},
but these phase transitions are not strong so a singlet scalar is
usually used to enhance the phase transition
\cite{Profumo:2007wc,Damgaard:2013kva,Ashoorioon:2009nf,OConnell:2006rsp,Gonderinger:2012rd,Profumo:2010kp,Gonderinger:2009jp,Barger:2008jx,
Cheung:2012nb,Barger:2007im,Cline:2013gha,Burgess:2000yq,Kakizaki:2015wua,Enqvist:2014zqa,Chala:2018ari,Noble:2007kk,Katz:2014bha,Espinosa:1993bs,Alanne:2014bra,Cline:2012hg,Beniwal:2017eik,Curtin:2014jma,Chiang:2018gsn,Dev:2019njv,Ghorbani:2018yfr,Ghorbani:2020xqv,Espinosa:2011ax,Espinosa:2007qk,
Kurup:2017dzf,Alves:2018jsw,Athron:2023xlk}, and also higher order
operators with a singlet extension are known to further enhance
the strength of the electroweak phase transition, making it strong
first order \cite{Oikonomou:2024jms}. As we show, in the present
context, the mirror world experiences a strong first order
electroweak phase transition at much higher temperatures than the
ordinary SM, and the stochastic signal of such a transition can be
detected by LISA, the BBO and DECIGO.

\section{High Scale Mirror Extended SM and its Overall Abundance}

\begin{figure}[htbp]
    \centering
    \includegraphics[width=0.6\textwidth]{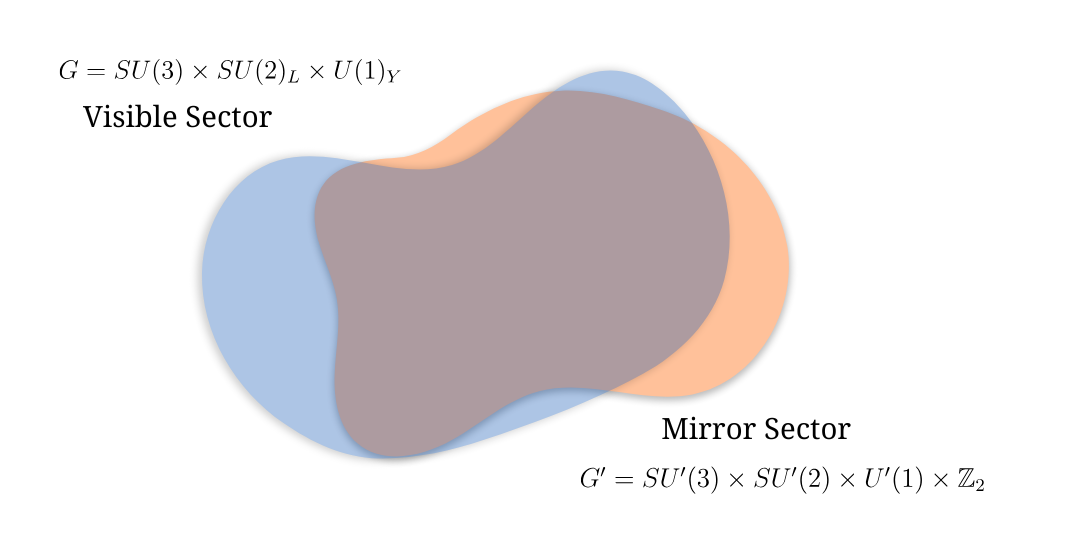}
    \caption{Schematic representation of the two ordinary SM and mirror SM sectors of the Universe alongside with the gauge group of each sector.}
    \label{fig:sectors}
\end{figure}

In the theoretical framework we consider in this world, the high
scale mirror SM is essentially a copy of the SM, with a singlet
extension coupled with the mirror Higgs sector, via renormalizable
and non-renormalizable terms. This extension is introduced via a
real singlet scalar field $\phi$ provided with a $\mathbb{Z}_2$
unbroken discrete symmetry \cite{Wang:2016uaz}, this symmetry
attributed to the fact that under the transformation $\phi$
$\rightarrow$ $-\phi$ every field remains unchanged, for more
details see section III. Subsequently the Universe is comprised of
ordinary particles with gauge group $ G=SU(3) \times SU(2)_L
\times U(1)_Y$ and mirror particles with gauge group $ G_M=SU(3)_M
\times SU(2)_M \times U(1)_M\times \mathbb{Z}_2$ (see
Fig.~\ref{fig:sectors}), the subscript ``M'' stands for the mirror
world. Eventually the total gauge group is described by $ G
\otimes G_M$ \cite{Oikonomou:2026ugb}. This paper states that part
of the DM of our Universe may be comprised by particles described
by this high scale extended SM. The scheme of this paper is based
on the fact that the two sectors do not have any elementary
interactions and interact only gravitationally with one another,
so we have a sort of hidden mirror DM model.
However the high scale mirror DM neutrinos, electrons and photons
contribute to the Big Bang Nucleosynthesis (BBN) of the real world
\cite{Berezhiani:2000gw}. This contribution leads to an effective
number of extra neutrino species in the bound of BBN, $\Delta
N_{\nu}$ =6.14. The current value of effective number of extra
neutrino species according to CMB/ BBN and PLANCK data is $\Delta
N_{\nu}$ = 0.35 $\pm$ 0.16 \cite{Nollett:2014lwa}, in the present
study the value $\Delta N_{\nu}$ \textless 0.4 is considered.
Therefore the density of mirror particles in the early Universe
should be degraded. This modification is plausible taking into
account two factors.
\begin{itemize}
    \item Firstly, the two sectors are not into thermal equilibrium, the
mirror sector has a lower temperature compering to the real
sector, $T' \textless T$
\cite{Silagadze:2008fa,Berezhiani:2000gw,Berezhiani:2003wj}. In
consideration of $\Delta N_{\nu}=6.14 \left( \frac{T'}{T}
\right)^4$ \cite{Ciarcelluti:2010zz,Oikonomou:2026ugb}, with $T'$
being the temperature of mirror world and $T$ the SM equilibrium
temperature, and the limit  $\Delta N_{\nu}$ \textless 0.4 it is
implied that $T' \textless 0.5 T$ \cite{Berezhiani:2000gw}.
    \item Secondly, the two sectors should interact weakly with one another,
this is automatically fulfilled considering only gravitational
interaction.
\end{itemize}
Hence, assuming that $T'= 0.5 \cdot T$, we compute the density
parameter of the mirror particles in the Universe $\Omega_{MSM}$
\footnote{In the context of this paper $\Omega_{MSM} \equiv
\Omega_{B}'$, since it is adopted that part of the dark matter can
be sourced from a mirror high scale SM.} with respect to density
parameter of ordinary standard model particles $\Omega_{SM}$
\cite{Berezhiani:2000gw},
\begin{equation}\label{energy density}
    \frac{\Omega_{MSM}}{\Omega_{SM}}=x^3 \cdot D^{-K(x)}
\end{equation}
with $K(x)=\frac{1-x^2}{\sqrt{1+x^4}}$, $x=\frac{T'}{T}$ and D
signifies the coefficient of the terms $\thicksim$  $T'^2$, see
section III. It is worth mentioning that in our Universe the total
density parameter is given by $\Omega_{total}$ =
$\Omega_{Baryons}$ + $\Omega_{\Lambda}$ + $\Omega_{DM}$ $\approx$
1, with the current values of density parameters being
$\Omega_{\Lambda}$=0.685,  $\Omega_{Baryons}$=0.0493 and
$\Omega_{DM}$=0.265,\cite{Planck:2018vyg}. Therefore $
\Omega_{MSM}=0.0493\cdot x^3 \cdot D^{-K(x)}$. Another interesting
topic to discuss is whether in the mirror Universe atoms could
form, namely, to examine the equation of binding energy in the
mirror sector. It is necessary to define the established mechanism
for calculating the binding energy and the recombination
temperature, $T_{rec}$ based on the parameters of the standard
model. The binding energy of the hydrogen atom is given by this
equation: \begin{equation} \label{binenergy}
    E_B=\frac{m_e \cdot a^2}{2}
\end{equation}
where the mass of the electron expressed as $m_e=\frac{y_e \cdot
u}{\sqrt{2}}$ and the fine structure is given by,
\begin{equation}
\label{finstr}
    a=\frac{g\cdot \tilde{g}'}{4\pi(g^2+\tilde{g}^2)}
\end{equation}
The term ``$u$'' is used here to refer to the scale of the
particle model, whereas the $y_e$ refers to the Yukawa coupling
constant of the electron. In the standard model these values are
defined as a=$\frac{1}{137}$, $y_e=2.94\cdot 10^{-6}$, $g$=0.653,
$\tilde{g}$=0.35 and the scale of SM $u=246$ GeV. Therefore the
$E_B \sim 13.6 \ eV$. Based on the binding energy the
recombination temperature, $T_{rec}$, can be deduced as follows
$\frac{1 - X_e^{\text{eq}}}{X_e^{\text{eq}}} =
\frac{4\sqrt{2}\,\zeta(3)}{\sqrt{\pi}}\, \eta
\left(\frac{T}{m_e}\right)^{3/2} e^{E_{\text{binding}}/T}$, where
$X_e$ is defined as the ionization fraction, $\zeta{(3)}$ is the
Riemann function and $\eta$ is defined as the baryon to photon
ratio $\eta=\frac{n_{B}}{n_{\gamma}}=\frac{n_{b}-\overline{n_{b}}
}{n_{\gamma}} \approx 6.1\cdot 10^{-10}$, thus $T_{rec}$=0.31 eV
for the visible sector of the Universe.
\cite{Oikonomou:2026ugb,Reina:2012fs}. Subsequently, the binding
energy and the recombination temperature are calculated using only
the couplings constants and the fine structure of a standard
model. This is crucial for this paper, since, phenomenologically,
it can be shown that the recombination era happens earlier in the
mirror sector of the Universe. However, it is important to note
that the baryon asymmetry  $\eta$ is larger in the mirror sector
compared to the standard model of the visible sector
\cite{Berezhiani:2000gw}.

\section{High Scale Mirror SM and the High Temperature Electroweak Phase Transition}

In this section a more detailed  theoretical framework for the
high scale mirror SM potential with the singlet extension is
presented. This model contains all the mirror particles but with a
higher scale $u$ comparing to the SM, which is $u_{SM}= 246 \
GeV$. The scale of this mirror SM model will be in the order of $u
\thicksim O(10^3)\ GeV $. In this section, a general formalism for
the finite temperature potential will be analyzed, as well as the
symmetry of the singlet extension. In addition, throughout this
work we adopt natural units, $\hbar = c = k_B = 1$, with all
dimensionful quantities expressed in GeV unless stated otherwise.

\subsection{Generic Formalism for the Effective Potential of the Mirror SM with Temperature Dependence}

First and foremost, as it was mentioned in section II, the SM of
the mirror sector is readjusted by a singlet extension which is
assumed to be active in the scale $M=15-50$TeV. For this singlet
field $\phi$ (quantities that refer to it will have the subscript
``s'') the effective potential contains only even powers of $\phi$
which can be understood by the fact that this field is equipped
with an unbroken discrete symmetry $\mathbb{Z}_2$, $\phi
\rightarrow -\phi$. Therefore the total gauge group of the mirror
sector is $G'=SU'(3) \times SU'(2) \times U'(1)\times
\mathbb{Z}_2$. The symmetry of the gauge group prevents the
particle of the singlet field from decaying into lighter
particles. Despite the fact that the singlet is stable under the
$\mathbb{Z}_2$, the mirror singlet field $\phi$ is coupled to the
Higgs. As a consequence the singlet can still be pair-produced or
pair-annihilated via the portal coupling $\lambda_{HS}$, and
eventually contribute to the mirror effective potential and the
mirror Higgs mass, $m_H'$, through mixing and loop contributions.
The coupling of the singlet field with the mirror Higgs and mirror
goldstone terms is apparent in the mirror SM potential written
below \eqref{eq:tree_potential}. The parameters that describe the
potential apart from those of mirror SM are  $
\mu_{H'},\lambda_{H'} ,\mu_S, \lambda_S,\lambda_{HS},\lambda$ and
the energy scale of the non-renormalizable dimension six operators
of the singlet extension, $M$. In this context the $
\mu_{'H},\lambda_{H'}$ are the mirror SM Higgs mass parameters,
while analogously the $\mu_S, \lambda_S$ have the same role for
the singlet field. Moreover, the parameter $\lambda_{HS}$ is
crucial since it characterizes the strength of the Higgs-singlet
coupling sectors of the Universe, therefore alters the electroweak
symmetry breaking of the mirror Higgs sector as shown in section
III. In the present paper a unitary gauge is assumed, thus the
tree level mirror singlet extended effective potential density is
expressed as follows,
\begin{equation}\label{eq:tree_potential}
    V_0(h', \phi') = -\frac{\mu_{H'}^2}{2}h'^2 + \frac{\lambda_{H'}}{4}h'^4 - \frac{\mu_S^2}{2}\phi'^2 + \frac{\lambda_S}{4}\phi'^4 + \frac{\lambda_{HS}}{2}h'^2\phi'^2 +
    \frac{\lambda}{2M^2}h'^2\phi'^4\, ,
\end{equation}
where we also included dimension six non-renormalizable operators
active at a high scale $M$. It is worth mentioning that the scale
of the potential is described as $u'=
\frac{\mu_{H'}}{\sqrt{\lambda_{H'}}}$ written in terms of the
parameters. In addition as it was mentioned before the term
$\frac{\lambda}{2M^2}h'^2\phi'^4$ is a non-renormalizable
dimension 6 operator which is suppressed by the energy scale of
$M$, thus this term contributes only by a small correction to the
potential at energy scale less than $M$ and it becomes
non-negligible in the thermal plasma described by energy scales
close to the value of $M$. For this study the value $M=15$TeV is
adopted. From equation \eqref{eq:tree_potential} is possible to
arrive at the definition of the effective mass of the particles,
taking into account only the mirror gauge bosons, the mirror top
quark, the mirror Higgs and mirror goldstone bosons. This is true,
since only these particles are deemed dominant in the one-loop
finite temperature effective potential
\cite{Oikonomou:2024jms,Espinosa:1993bs}. Thus, the resulting
expressions for the mass of the mirror particles are,
\begin{equation}
m_h'^2(h', \phi') = -\mu_{H'}^2 + 3\lambda_{H'} h'^2 + \lambda_{HS}\phi'^2 + \frac{\lambda}{M^2}\phi'^4, \label{eq:mass_h}
\end{equation}

\begin{equation}
     m_\chi'^2(h', \phi') = -\mu_{H'}^2 + \lambda_{H'} h'^2 + \lambda_{HS}\phi'^2 + \frac{\lambda}{M^2}\phi'^4, \label{eq:mass_chi}
\end{equation}

\begin{equation}
    m_S'^2(h', \phi') = -\mu_S^2 + 3\lambda_S \phi'^2 + \lambda_{HS} h'^2 + \frac{6\lambda}{M^2}h'^2\phi'^2, \label{eq:mass_S}
\end{equation}

\noindent
\begin{minipage}{0.33\textwidth}
    \begin{equation}
        m_W'^2(h') = \frac{g'^2}{4}h'^2 \label{eq:mass_W}
    \end{equation}
\end{minipage}%
\begin{minipage}{0.33\textwidth}
    \begin{equation}
        m_Z'^2(h') = \frac{g'^2 + \tilde{g}'^2}{4}h'^2 \label{eq:mass_Z}
    \end{equation}
\end{minipage}%
\begin{minipage}{0.33\textwidth}
    \begin{equation}
        m_{t'}^2(h') = \frac{y_{t'}^2}{2}h'^2 \label{eq:mass_t}
    \end{equation}
\end{minipage}
The term $m_\chi'$ is used here to refer to the mirror goldstone
bosons, while the term $m_S'$ is defined as the mass of the mirror
singlet particle. It is important to note that the $g',
\tilde{g}'$ and $y_{t'}$ denotes the coupling constants of  $
SU'(2)_L, \ U'(1)_Y$ and mirror coupling of the top quark,
accordingly. In this paper, we adopt the values $g'=0.653$,
$\tilde{g}'=0.357$ and $y_t'=0.995$ for the mirror sector,
therefore from \eqref{finstr} the value of mirror fine-structure
is calculated as $\alpha'=0.03349$. In order to develop the
correct expression for the finite temperature potential, three
more elements are needed. Namely it is necessary to add zero
temperature one-loop corrections\footnote{The $\overline{MS}$
renormalization is applied here in order to prevent UV
divergences} in the tree level potential $ V_0(h', \phi')$ and
afterwards introduce finite temperature corrections with an extra
term, $V'^i_{ring}$, for the resummation of the leading infrared
divergences arising from the bosonic zero Matsubara modes.
Generally speaking the potential is written as,
\begin{equation} \label{Veff}
    V'_{\text{eff}}(h', \phi', T') = V'_0(h', \phi') + \sum_i \left[ V'^i_1 \left(m'^2_i(h', \phi')\right) + V'^i_{T'} \left(m'^2_i(h', \phi'), T'\right) + V'^i_{ring} \left(m'^2_i(h', \phi'), T'\right) \right]
\end{equation}
where $i = \{h', \chi',\phi, W', Z',t', \gamma'\}$. This potential
is written accordingly to Arnold and Espinosa
\cite{Espinosa:1993bs} scheme in order to treat the infrared
divergencies caused by zero Matsubara modes. Elaborating on that,
the term $V_{ring}$ is given by the expression,
\begin{equation}\label{ringterm}
    V'^i_{ring} \left(m'^2_i(h', \phi'), T'\right) = \frac{\bar{n}'_i T'}{12\pi} \left[ m'^3_i(h', \phi') - \left( m'^2_i(h', \phi') + \Pi'_i(T') \right)^{3/2} \right] ,
\end{equation}
where once again  $i = \{h', \chi',\phi, W', Z', \gamma'\}$ and
$\bar{n_i}$ is defined as the degree of freedom of each particle,
$\bar{n_i}=\{1,3,1,2,1,1\}$ accordingly. It is important to note
that the term $ \Pi'_i(T')$, appearing in equation
\eqref{ringterm}, is called thermal (Debye) self-energy of mirror
gauge bosons-scalar fields and is essentially the leading order in
the one-loop correction to the mass term of the particles due to
the interaction with surrounding thermal plasma. The self-energy
of each field can be found in appendix A. The term $ M_i'^2
=m'^2_i(h', \phi') + \Pi'_i(T') $ is called thermal mass and it is
used in $ V'^i_{ring}$ potential. As a result of this formalism
the effective potential in the renormalization scale $\mu_R=2\cdot
m_{t'}$ is given by the equation,
\begin{equation}
\label{eq:V_eff_mirror}
\begin{split}
    V'_{\text{eff}}(h', \phi', T') &= - \frac{\mu'^2_H}{2}h'^2 + \frac{\lambda'_H}{4}h'^4 - \frac{\mu'^2_S}{2}\phi'^2 + \frac{\lambda'_S}{4}\phi'^4 + \frac{\lambda'_{HS}}{2}h'^2\phi'^2 + \lambda'\frac{h'^2\phi'^4}{2M'^2} \\[1ex]
    &\quad + \sum_i \frac{n'_i m'^4_i(h', \phi')}{64\pi^2} \left[ \ln\left(\frac{m'^2_i(h', \phi')}{\mu'^2_R}\right) - C'_i \right] - \frac{n'_t m'^4_t(h', \phi')}{64\pi^2} \left[ \ln\left(\frac{m'^2_t(h', \phi')}{\mu'^2_R}\right) - C'_t \right] \\[1ex]
    &\quad + \sum_i \frac{n'_i T'^4}{2\pi^2} J_B \left( \frac{m'^2_i(h', \phi')}{T'^2} \right) - \frac{n'_t T'^4}{2\pi^2} J_F \left( \frac{m'^2_t(h')}{T'^2} \right) \\[1ex]
    &\quad + \sum_i \frac{\bar{n}'_i T'}{12\pi} \left[ m'^3_i(h', \phi') - \left( M'^2_i(h', \phi', T') \right)^{3/2} \right] ,
\end{split}
\end{equation}

Then by performing a high temperature expansion into the thermal
functions,
$$J_{B/F}(y^2)=\int_0^\infty dx \, x^2 \ln \left[ 1 \mp
\exp \left( -\sqrt{x^2 + y^2} \right) \right]\, ,$$ while
substituting the degrees of freedom, $\bar{n_i}$, for each
particle and the parameter $C'_{i}=3/2$ for scalars-fermions and
$C'_{i}=5/2$ for gauge bosons, we end up with the final expression
of mirror effective potential,
\begin{equation}
\label{final V_HT_mirror}
\begin{split}
    V^{'}_{\text{eff}}(h', \phi', T') &= -\frac{\mu'^2_H}{2}h'^2 + \frac{\lambda'_H}{4}h'^4 - \frac{\mu'^2_S}{2}\phi'^2 + \frac{\lambda'_S}{4}\phi'^4 + \frac{\lambda'_{HS}}{2}h'^2\phi'^2 + \frac{\lambda'}{2M'^2}h'^2\phi'^4 \\[1ex]
    &\quad + \frac{m'^2_h(h', \phi')}{24}T'^2 - \frac{T'}{12\pi} [M'^2_h(h', \phi', T')]^{3/2} + \frac{m'^4_h(h', \phi')}{64\pi^2} \left[ \ln\left(\frac{a_b T'^2}{\mu'^2_R}\right) - \frac{3}{2} \right] \\[1ex]
    &\quad + \frac{3m'^2_\chi(h', \phi')}{24}T'^2 - \frac{3T'}{12\pi} [M'^2_\chi(h', \phi', T')]^{3/2} + \frac{3m'^4_\chi(h', \phi')}{64\pi^2} \left[ \ln\left(\frac{a_b T'^2}{\mu'^2_R}\right) - \frac{3}{2} \right] \\[1ex]
    &\quad + \frac{m'^2_\phi(h', \phi')}{24}T'^2 - \frac{T'}{12\pi} [M'^2_\phi(h', \phi', T')]^{3/2} + \frac{m'^4_\phi(h', \phi')}{64\pi^2} \left[ \ln\left(\frac{a_b T'^2}{\mu'^2_R}\right) - \frac{3}{2} \right] \\[1ex]
    &\quad + \frac{6m'^2_W(h')}{24}T'^2 - \frac{4T'}{12\pi} m'^3_W(h') - \frac{2T'}{12\pi} [M'^2_{W_L}(h', T')]^{3/2} + \frac{6m'^4_W(h')}{64\pi^2} \left[ \ln\left(\frac{a_b T'^2}{\mu'^2_R}\right) - \frac{5}{6} \right] \\[1ex]
    &\quad + \frac{3m'^2_Z(h')}{24}T'^2 - \frac{2T'}{12\pi} m'^3_Z(h') - \frac{T'}{12\pi} [M'^2_{Z_L}(h', T')]^{3/2} + \frac{3m'^4_Z(h')}{64\pi^2} \left[ \ln\left(\frac{a_b T'^2}{\mu'^2_R}\right) - \frac{5}{6} \right] \\[1ex]
    &\quad + \frac{12m'^2_t(h')}{48}T'^2 - \frac{12m'^4_t(h')}{64\pi^2} \left[ \ln\left(\frac{a_f T'^2}{\mu'^2_R}\right) - \frac{3}{2} \right] - \frac{T'}{12\pi} [M'^2_{\gamma_L}(h', T')]^{3/2} ,
\end{split}
\end{equation}
with $a_b=223.0993$ and $a_f = 13.943$ \cite{Profumo:2007wc}.

\begin{figure}[htbp]
    \centering

    \begin{subfigure}[b]{0.34\textwidth}
        \centering
        \includegraphics[width=\linewidth]{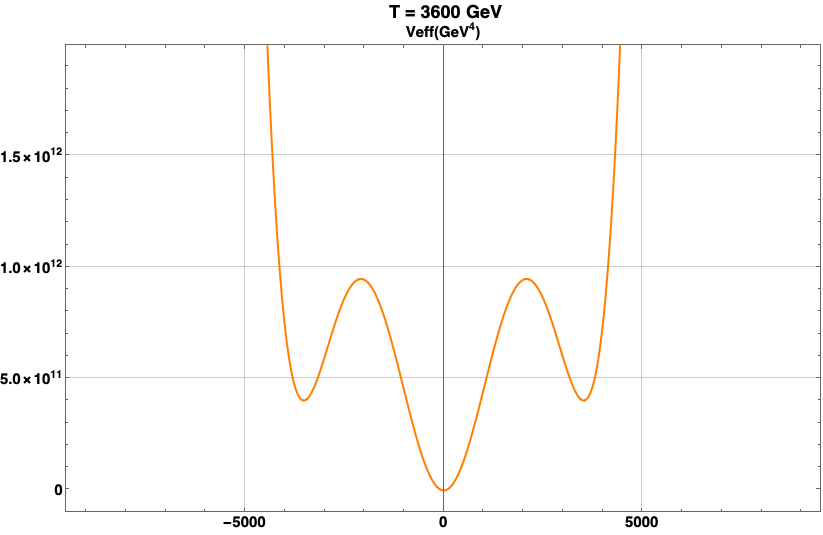}
        \label{fig:file1}
    \end{subfigure}%
    \hfill%
    \begin{subfigure}[b]{0.32\textwidth}
        \centering
        \includegraphics[width=\linewidth]{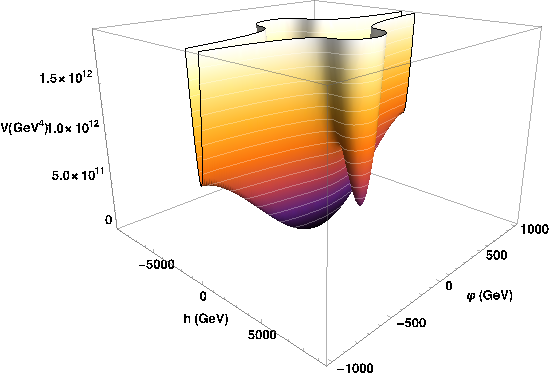}
        \label{fig:file2}
    \end{subfigure}%
    \hfill%
    \begin{subfigure}[b]{0.3\textwidth}
        \centering
        \includegraphics[width=\linewidth]{file3.png}

        \label{fig:file3}
    \end{subfigure}

    \caption{ \textbf{Left}: The effective potential of the high scale mirror SM at temperature
$T= 3600\ GeV$. Note that $T= 3600\ GeV$ is above $T_c$, since the
point ($0,0$) still remains global minimum in the Higgs sector.
    \textbf{Middle}: The effective potential in 3 dimensions, as it can be seen the
symmetry along side the $\phi$ remains unbroken.
\textbf{Right}:Two-dimensions density plot of h over $\phi$
Behavior of the effective potential in the field space. It is
important to note that the symmetry of singlet scalar field
remains unbroken alongside the $\phi$  direction.}

    \label{fig:effective-potential}
\end{figure}

In Fig. \ref{fig:effective-potential} an eminent example of the
finite temperature effective potential is demonstrated, depending
on the choice of the free parameters. In the case at hand, the
scale of the mirror sector is $u'=5.1 \ TeV$ and the values
assigned to the parameters are set to $\mu'_H = 813.272 \ GeV$,
$\mu'_s =47.766 \ GeV$, $\lambda'_S=1.03436 \cdot 10^{-5}$,
$\lambda'_{HS}=5.99787$ and $\lambda=4.35164 \cdot10^{-5}$.
Now that the formalism of the effective potential is developed, it
is vital to deduce the definition of the parameter $D$ mentioned
in the Eq. \eqref{energy density}. In the literature
\cite{Oikonomou:2026ugb,Oikonomou:2024geq} the term $D$ is defined
as the coefficient of the quadratic thermal contribution
$\thicksim T^{'2}$ to the Higgs field power $h'^2$. Therefore by
performing a computational analysis to determine the coefficient
of $T'^2$ considering the terms $V^{'}_{\text{eff}}(h', \phi', T')
\thicksim D(T'^2-T'^2_0) \cdot h'^{2} $,  we arrive at the
definition of $D$ as,
\begin{equation}\label{eq:D_parameter}
    D=\frac{3}{32}g'^{2} + \frac{1}{32} \tilde{g}'^{2}+ \frac{1}{8} y_t'^2 + \frac{\lambda_H'}{4} + \frac{\lambda_{HS}'}{24}
\end{equation}
The significant modification in the equation given and the one
mentioned in the literature, is the term
$\frac{\lambda_{HS}'}{24}$. This term is crucial since it defines
the energy density, $\Omega_{MSM}$, of the mirror sector of the
Universe which can be comprehended via Eq. \eqref{energy density}.
It is noteworthy that even though the symmetry of the singlet
field is not broken, therefore the vacuum expectation value of the
field is equal to zero $\langle \phi \rangle=0 $, the mirror
Higgs-Singlet portal coupling constant $\lambda_{HS}'$ performs a
central role in establishing the values of the physical parameters
of the effective theory.

Before proceeding to the analytical evaluation of some high scale
mirror dark matter models it is important to clarify further how
the symmetry along the $\phi$ direction is ensured during the
procedure of cosmological phase transition. First, the singlet
direction is locally stable at the mirror electroweak vacuum,
since $m_S'^2(u',0) = -\mu_S^2 + \lambda_{HS}\,u'^2 > 0$ for all
parameter points considered \footnote{the $m_S'^2(u',0) = -\mu_S^2
+ \lambda_{HS}\,u'^2$ formula arises from \eqref{eq:mass_S} at the
locus of point $(h,\phi)=(u,0)$.}. Second, the mirror electroweak
minimum is the global one: numerically is verified that
$V'_{\text{eff}}(u',0,T') < V'_{\text{eff}}(0,w',T')$, where $w'$
denotes the would-be minimum along the singlet direction; at tree
level this is guaranteed by the sufficient condition
$\lambda_{HS}^{\text{eff}} > \sqrt{\lambda_{H'}\lambda_S}$, with
$\lambda_{HS}^{\text{eff}}$ including the shift induced by the
dimension-six operator, which simultaneously excludes the
existence of a mixed minimum with $\langle h'\rangle,
\langle\phi'\rangle \neq 0$. Third, at finite temperature the
positive thermal self-energy $\Pi_S(T') > 0$ further stabilizes
the $\phi' = 0$ direction, so the $\mathbb{Z}_2$ is restored
rather than broken at high $T'$, as confirmed numerically in
Fig.~\ref{fig:effective-potential}. Since the discrete symmetry is
never spontaneously broken, the model is automatically free of the
cosmological domain-wall problem, while the unbroken
$\mathbb{Z}_2$ guarantees the stability of the singlet.

\subsection{Singlet-extended Mirror Electroweak Phase Transition}

All fascinating physical phenomena emerge from the spontaneous
symmetry breaking in the mirror Higgs sector of the effective
potential, which breaks at a finite temperature in the early
Universe. The expectation value of the mirror Higgs field changed
from $0 \rightarrow u$, where $u$ is the scale of the model, since
the point in the $h$ axis that minimizes the potential shifts from
$h=0\rightarrow h=u$. The value of $u$ provides mass to the
particle as implied in Eqs. \eqref{eq:mass_h},
\eqref{eq:mass_chi}, \eqref{eq:mass_S}, \eqref{eq:mass_W},
\eqref{eq:mass_Z} and \eqref{eq:mass_t}. We shall dub the shift in
the expectation value of mirror Higgs field as mirror electroweak
phase transition and it be described as the transformation in the
gauge group as $SU(2)_L \times U(1)_Y \xrightarrow{\langle h
\rangle \neq 0} U(1)_{EM}$. In this study on the case of mirror
SM, equipped by a $\mathbb{Z}_2$ unbroken symmetry, the shift in
the gauge group is presented as $SU(2)_L \times U(1)_Y \times
\mathbb{Z}_2 \xrightarrow{\langle h \rangle \neq 0} U(1)_{EM}
\times \mathbb{Z}_2$. As discussed above a first order phase
transition in the mirror potential presents an immense interest
physically, since it can lead to the production of stochastic
gravitational waves, therefore contribute as gravitational wave
background in the low frequency section of the gravitational wave
spectrum, this is further reviewed in the next section. In the
rest of this section, we present three high scale mirror models
with different high scale vacuum expectation values $u$.


\subsubsection{Model I: Scale $u  \thicksim 6000 \ GeV$}

In this model, the mirror Yukawa couplings have the value
mentioned above, whereas the scale is $u'_1=6125.118 \ GeV$ and
the parameters  $\mu'_H = 25.114 \ GeV$, $\mu'_s =100.183  \ GeV$,
$\lambda'_S=0.0190587 $, $\lambda'_{HS}=6.01427$.  For this choice
of parameters the mass of the mirror particles is quoted in Table
\ref{tab:masses-model1}, using equations \eqref{eq:mass_h},
\eqref{eq:mass_W}, \eqref{eq:mass_Z} and \eqref{eq:mass_t}.

\begin{table}[htbp]
    \centering
    \renewcommand{\arraystretch}{1.3}
    \setlength{\tabcolsep}{1cm}
    \begin{tabular}{|c|c|}
        \hline
        \multicolumn{2}{|c|}{Mirror particle mass} \\
        \hline
        Particle & Value \\
        \hline
        $H'$ & $m_H'=35.52\ GeV$ \\
        \hline
        $S'$ & $m_S'=15020.9 \  GeV$ \\
        \hline
        $W'$ & $m_W'=1999.85 \ GeV$ \\
        \hline
        $Z'$ & $m_Z'=2279.21 \ GeV$ \\
        \hline
        $t'$ & $m_t'=4309.46 \  GeV$ \\
        \hline
    \end{tabular}
    \caption{Mirror SM Masses: Model 1}
    \label{tab:masses-model1}
\end{table}
In addition to that, by assuming the Yukawa coupling of the mirror
electron is equal to the one of SM, i.e. $y_e = 2.94 \cdot
10^{-6}$, the mass of the mirror electron is equal to $m_e'
\thicksim 12.73$ MeV, which is heavier due to the high scale
model. From Eqs. \eqref{binenergy} and \eqref{finstr} we arrive at
the value of mirror binding energy $E_B'\thicksim 7.141 \ keV$,
which is significantly larger than the $E_B \thicksim13.6 \ eV$ of
the visible sector. Therefore, the conclusion that atoms form
quite earlier in the mirror sector is reached. As the binding
energy is larger, the atomic structure is more tightly bound
against the high-temperature plasma in the early mirror sector, in
other words the atoms resist against the ionization of the hot
plasma. Furthermore, the temperature of the mirror sector is less
than the one in the visible sector, and this further supports the
claim that atoms form much earlier in the mirror sector than in
the visible sector.

\begin{figure}[htbp]
    \centering

    \includegraphics[width=0.7\textwidth]{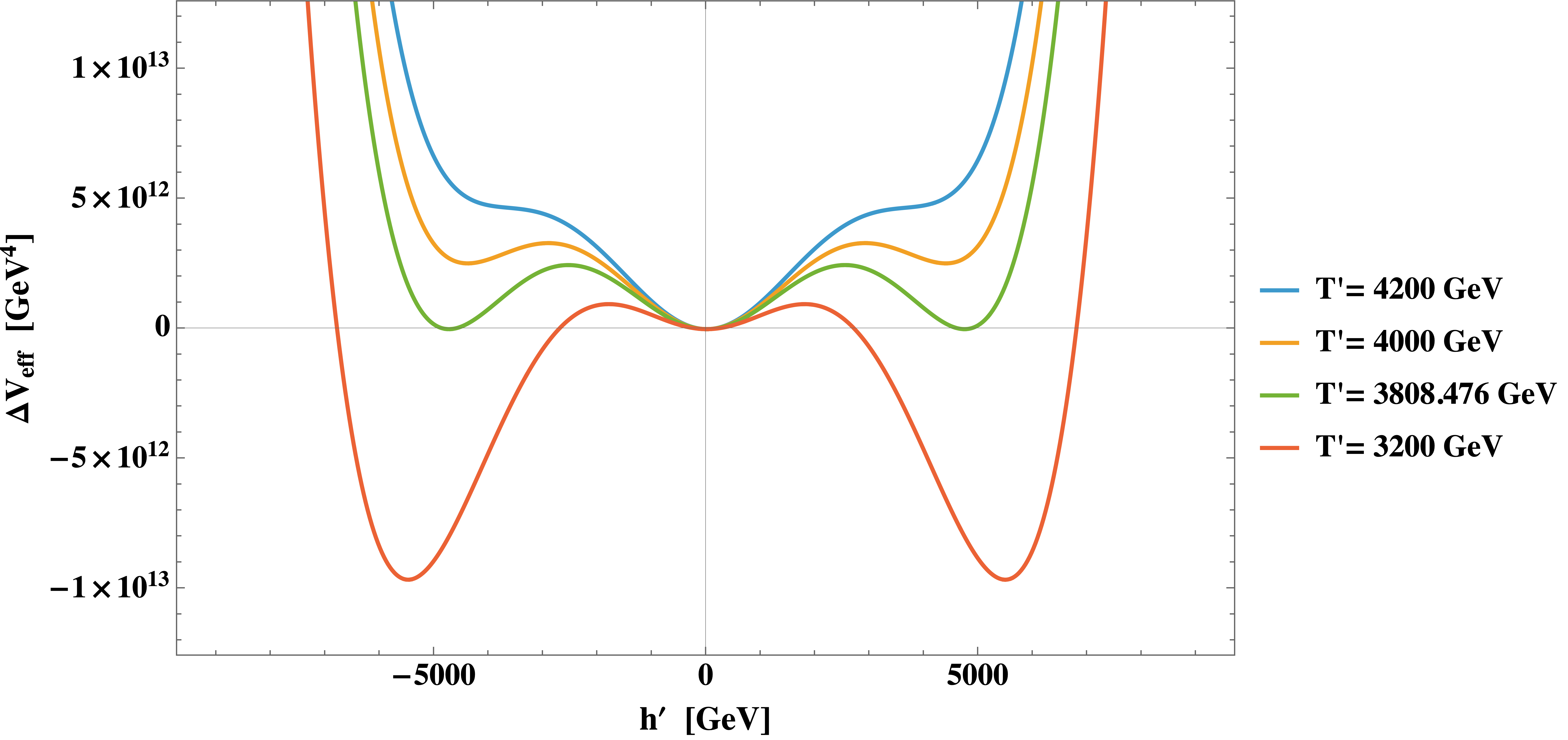}
    \caption{Behavior of the effective potential for the mirror SM with $u_1'$
in various temperatures near the critical temperature
$T_c\thicksim 3808.476$ GeV. This figures clearly shows a
first-order phase transition. The phase transition is considered
strong since $\frac{u_c'}{T_c'}=1.24$}
    \label{fig:model1}
\end{figure}

Returning to the behavior of the effective potential regarding the
temperature of the surrounding environment, it is clear that the
system undergoes a first order phase transition. The
finite-temperature effective potential exhibits a barrier between
the symmetric and broken minima, demonstrating that the
electroweak phase transition is of first order, with the critical
temperature being equal to $T'_c \thicksim 3808.476$ GeV, where
the minima are degenerate. More details can be deduced from
Fig.~\ref{fig:model1} \footnote{Note that the y-axis is defined as
$V^{'}_{\text{eff}}(h', \phi', T')-V^{'}_{\text{eff}}(0,0, T')$ in
order to situate the false vacuum at the start of the cartesian
system. This is physically allowed since introducing the constant
$C = V^{'}_{\text{eff}}(0,0, T')$ transforms the Lagrangian as
$\mathcal{L}' = \mathcal{L} + C$. Considering $\partial C /
\partial \Phi_i = 0$ for the fields $\Phi_i \in \{h', \phi'\}$,
the equations of motion remain completely unaffected.}

The transition does not complete at the critical temperature due
to the barrier that is formed between the two local minima of the
potential. The phase transition is a supercooled one with the
percolation temperature being $T_*'\thicksim 1142\ GeV$, where the
true vacuum $h' \ne 0 \ GeV$ is more energetically favored than
the false vacuum. In order to assess the strength of this
first-order phase transition, the sphaleron rate criterion will be
used, i.e. the value of
\cite{Ramsey-Musolf:2019lsf,Profumo:2007wc,Ahriche:2014jna},
\begin{equation}
    \frac{u'_c}{T_c'} \gtrsim \eta_k \   \text{and} \ \ \eta_k \simeq 0.6 -1.4
\end{equation}
In this model the ratio is computed at $\frac{u'_c}{T_c'} = 1.24$,
therefore this phase transition can be classified as a first order
strong phase transition. Based on past literature
\cite{Apreda:2001us,Schabinger:2005ei,Kusenko:2006rh,McDonald:1993ex,Chala:2018ari,Davoudiasl:2004be,Baldes:2016rqn,Noble:2007kk,Zhou:2020ojf,
Weir:2017wfa,Hindmarsh:2020hop,Han:2020ekm,Child:2012qg,Fairbairn:2013uta,LISACosmologyWorkingGroup:2022jok,Caprini:2015zlo,Huber:2015znp,
Delaunay:2007wb,Chung:2012vg,Barenboim:2012nh,Senaha:2020mop,Grojean:2006bp,Katz:2014bha,Alves:2018jsw,Athron:2023xlk}
a strong first order cosmological phase transition  produces
stochastic gravitational waves via the collision of bubble
containing regions of space with non-zero vacuum expectation value
of the mirror Higgs field, that will be studied in the next
section of this paper. The abundance of DM is computed, using
equations \eqref{energy density}, \eqref{eq:D_parameter} as
$\Omega_{B'} \thicksim0.012$. Therefore only a fraction of the DM
can be attributed to the high-scale mirror DM for this model.

\subsubsection{Model II: Scale $u  \thicksim 5000 \ GeV$}

\begin{figure}[htbp]
    \centering
    \includegraphics[width=0.7\textwidth]{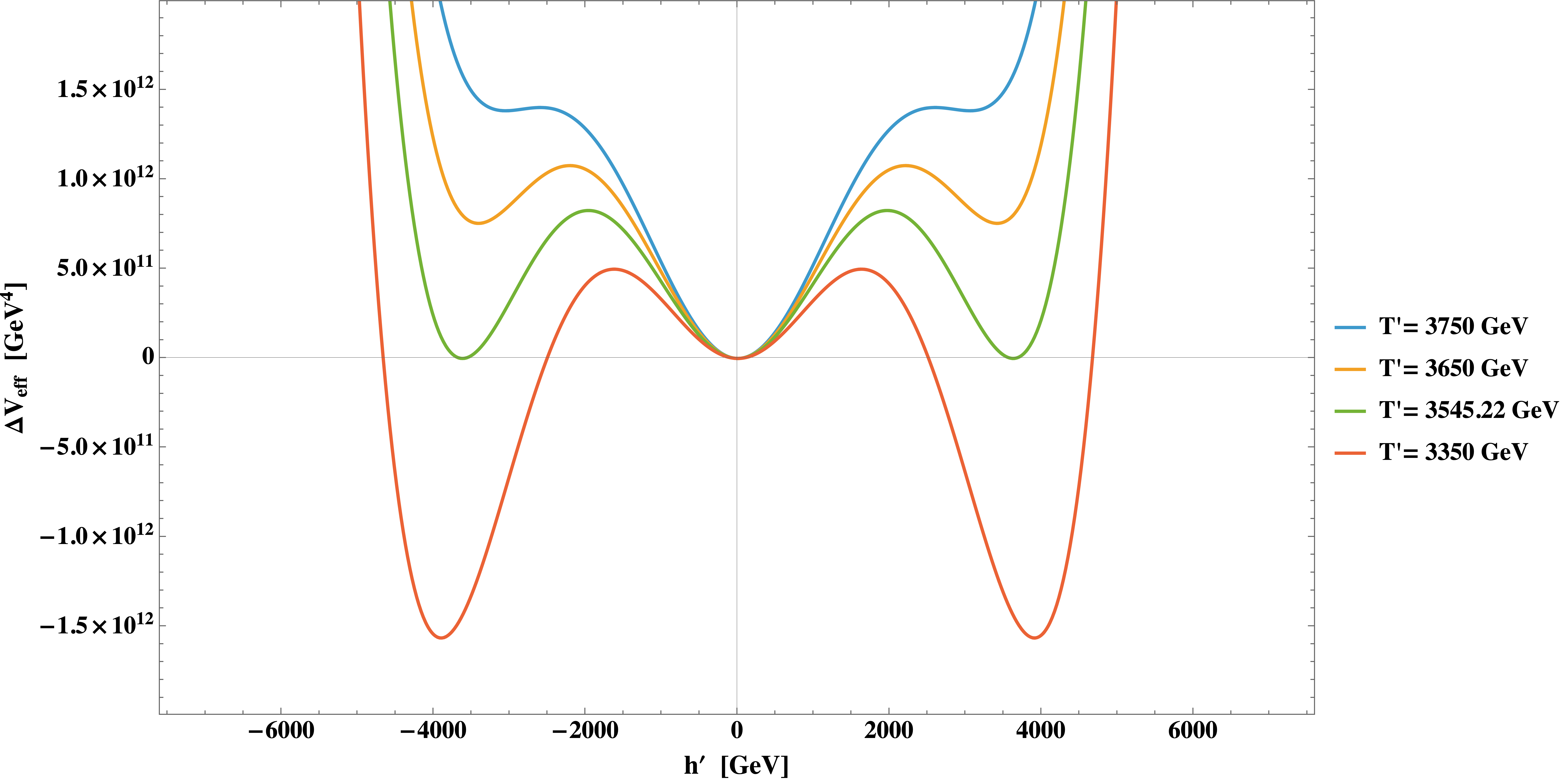}
    \caption{Behavior of the effective potential for the mirror SM with $u_2'$
in various temperatures near the critical temperature $T_c' =
3545.22$ GeV. This figure clearly shows a first-order phase
transition with the sphaleron rate criterion being considered, the
phase transition is deemed strong since $\frac{u_c'}{T_c'}=1.02$.}
    \label{fig:model2}
\end{figure}

Let us study another high scale model the parameter point $\mu'_H
= 813.272 \ GeV$, $\mu'_s =47.766  \ GeV$,
$\lambda'_S=1.03436\cdot 10^{-5} $, $\lambda'_{HS}=5.99787$ is
chosen and the scale of the finite temperature effective potential
is equal to  $u'_2=5052.421 \ GeV$. Once again the mass of the
mirror particles is determined for this choice of parameters using
equations \eqref{eq:mass_h}, \eqref{eq:mass_W}, \eqref{eq:mass_Z}
and \eqref{eq:mass_t}. The results are presented in Table
\ref{tab:masses-model2}.
\begin{table}[htbp]
    \centering
    \renewcommand{\arraystretch}{1.3}
    \setlength{\tabcolsep}{1cm}
    \begin{tabular}{|c|c|}
        \hline
        \multicolumn{2}{|c|}{Mirror particle mass} \\
        \hline
        Particle & Value \\
        \hline
        $H'$ & $m_H'=1150.14\ GeV$ \\
        \hline
         $S'$ & $m_S'=12373.6 \  GeV$ \\
        \hline
        $W'$ & $m_W'=1649.62 \ GeV$ \\
        \hline
        $Z'$ & $m_Z'=1880.05 \ GeV$ \\
        \hline
        $t'$ & $m_t'=3554.74 \  GeV$ \\
        \hline
    \end{tabular}
    \caption{Mirror SM Masses: Model 2}
    \label{tab:masses-model2}
\end{table}
We can now compute the mass of the mirror electron in order to
investigate whether mirror atoms can be formed. Based on the
assumption that the Yukawa coupling of the mirror electron is
essentially the same as the one of SM, namely $y_e = 2.94 \cdot
10^{-6}$, the mass of the mirror electron is determined as $m_e'
\thicksim 10.5$ MeV which is smaller than the mirror electron mass
of the model I due to the value of $u'_2$. In addition, the
binding energy in the mirror sector of hydrogen is determined
using Eq. \eqref{binenergy}, therefore $E_B'\thicksim 5.89  \ keV$
which is greater than the value of SM, thus atoms form earlier in
the mirror sector in this case. Moving on now to consider the
behavior of the finite temperature potential regarding the
temperature dependence we arrive at the conclusion that this model
also presents a typical first order phase transition with regard
to the temperature, which is demonstrated in
Fig.~\ref{fig:model2}.
The model presented above is a typical first-order phase
transition since a barrier is formed between the true and the
false vacuum of the Higgs sector. Therefore, the two phases
co-exist in the same physical space and the bubbles of the true
vacuum of Higgs sector nucleate and collide producing stochastic
gravitational waves. The critical temperature is determined at the
value $T_c' \thicksim 3545 \ GeV$. According to the behavior of
the finite temperature effective potential, the percolation
temperature, at which expanding bubbles of the true vacuum occupy
a sufficiently large fraction of space, is deemed quite lower than
$T_c'$ at the value of $T'_* \thicksim  886 \ GeV$, thus the phase
transition is a supercooled one. Once again by adopting the
sphaleron rate criterion, we obtain $\frac{u'_c}{T_c'} = 1.02$,
and thus we have the realization of a strong phase transition. In
addition, the abundance of the high scale SM is equal to
$\Omega_{B'} \thicksim0.012$, so in this case too, the mirror DM
can be a small portion of the total DM of the Universe.

\subsubsection{Model III: Scale $u  \thicksim 4000 \ GeV$}

In the third model for the high-scale finite temperature potential
we choose the values of the parameters as $\mu'_H = 96.21 \ GeV$,
$\mu'_s =184.649  \ GeV$, $\lambda'_S=0.0539572$,
$\lambda'_{HS}=6.02011$, the vacuum expectation value of the Higgs
is equal to $u'_3=3931.969 \ GeV$. Relying on these values we
determine the mass of the mirror electron and the binding energy
of the mirror hydrogen atom \eqref{binenergy}, \eqref{finstr} and
considering that $m_e'=\frac{y_e' \cdot u'_3}{\sqrt{2}}$, $m_e'
\thicksim 8.17\ MeV$ and we get $E_B'\thicksim 4.585  \ keV$,
which is greater than the value of SM once more. The mass of the
mirror particles of this model are displayed in Table
\ref{tab:masses-model3}.

\begin{figure}[htbp]
    \centering
    \includegraphics[width=0.7\textwidth]{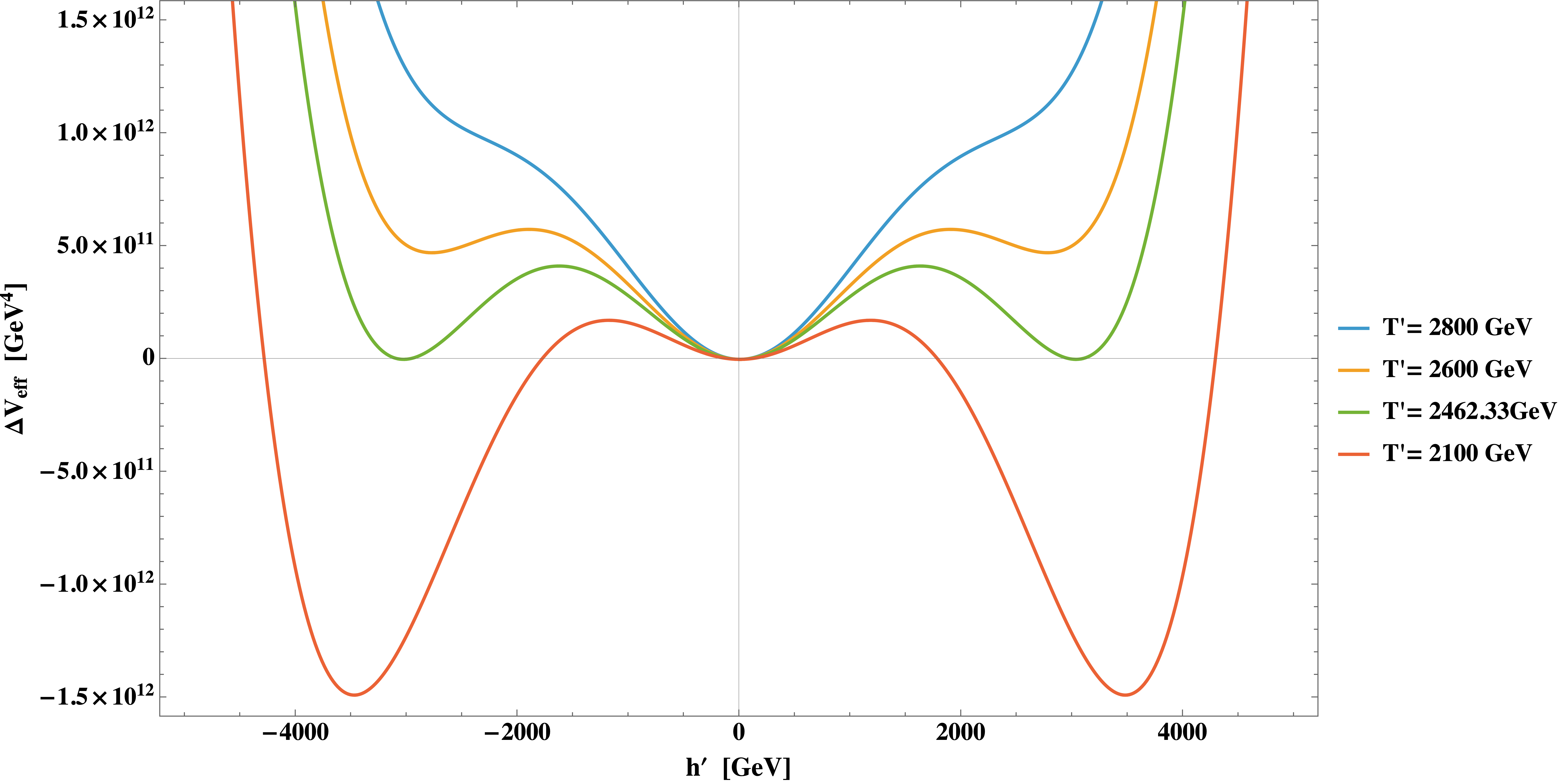}
    \caption{Model 3: Behavior of the effective potential for the mirror SM in
various temperatures near the critical temperature $T_c' = 2462.33
$ GeV. This figure clearly shows a first-order phase transition.
The sphaleron rate criterion is considered, therefore the phase
transition is considered strong since $\frac{u_c'}{T_c'}=1.23$.}
    \label{fig:model3}
\end{figure}

\begin{table}[htbp]
    \centering
    \renewcommand{\arraystretch}{1.3}
    \setlength{\tabcolsep}{1cm}
    \begin{tabular}{|c|c|}
        \hline
        \multicolumn{2}{|c|}{Mirror particle mass} \\
        \hline
        Particle & Value \\
        \hline
        $H'$ & $m_H'=136.06\ GeV$ \\
        \hline
         $S'$ & $m_S'=9645.68 \  GeV$ \\
        \hline
        $W'$ & $m_W'=1283.79 \ GeV$ \\
        \hline
        $Z'$ & $m_Z'=1463.12 \ GeV$ \\
        \hline
        $t'$ & $m_t'=2766.42 \  GeV$ \\
        \hline
    \end{tabular}
    \caption{Mirror SM Masses: Model 3}
    \label{tab:masses-model3}
\end{table}
Once more, we study the behavior of finite temperature potential
to determine whether the system experiences a first or a second
order phase transition. From Fig. \ref{fig:model3}, it is apparent
that in the case of model III, the phase transition is a first
order one, realized around the temperature 2500 GeV.
The critical temperature is determined as the temperature which
the two minima of the potential are the same, which is $T_c' =
2462.33 $ GeV. The transition from the false vacuum to the true
one starts and continues at $T_*'=738.698 \ GeV$, where the
expanding bubbles of true vacuum cover a large fraction of volume
in space. Moving on now to consider the strength of the phase
transition using the sphaleron rate criterion, it is calculated
that $\frac{u'_c}{T_c'} = 1.23$, thus this is a strong first order
phase transition, which once again can lead to the production of
gravitational waves. In addition, the abundance of the high scale
SM is equal to $\Omega_{B'} \thicksim0.012$.

This section began by describing the formalism of the finite
temperature effective potential at high scale $u$, while arguing
that a mirror extended SM is a viable candidate for DM. Despite
the fact that the energy density of mirror baryons is not
sufficient to explain the total DM of the visible sector, it still
comprises a fraction of the total DM of the Universe. In addition
the study of the electroweak phase transition in the dark sector
was developed, by elaborating on three different energy scales for
the vacuum expectation value of the Higgs field, regarding the
strength of the phase transition and it was deduced that all three
models realize a strong first order phase transition based on the
sphaleron rate criterion.


The section that follows moves on considering the effect of a
strong first order phase transition on the gravitational wave
spectrum. In particular the mechanism for producing GW by bubble
collision is reviewed, as well as the assessment of generated
gravitational waves from the three strong phase transition studied
in the present section, regarding the signal detectability of
gravitational waves takes place.

\section{SEARCHING FOR EFFECTS ON STOCHASTIC GRAVITATIONAL WAVES FROM THE DARK PHASE TRANSITIONS}

\subsection{Bubble Collision as a Method of Stochastic Gravitational Wave Production}

Before proceeding to examine the gravitational waves generated
from the three models mentioned in the previous section, it is
important to elaborate on certain physical aspects  regarding the
production of gravitational waves assuming bubble nucleation,
expansion and collision due to a first order phase transition,
namely the difference of the potential energy $\Delta V$.

\begin{figure}[htbp]
    \centering
    \includegraphics[width=0.7\textwidth]{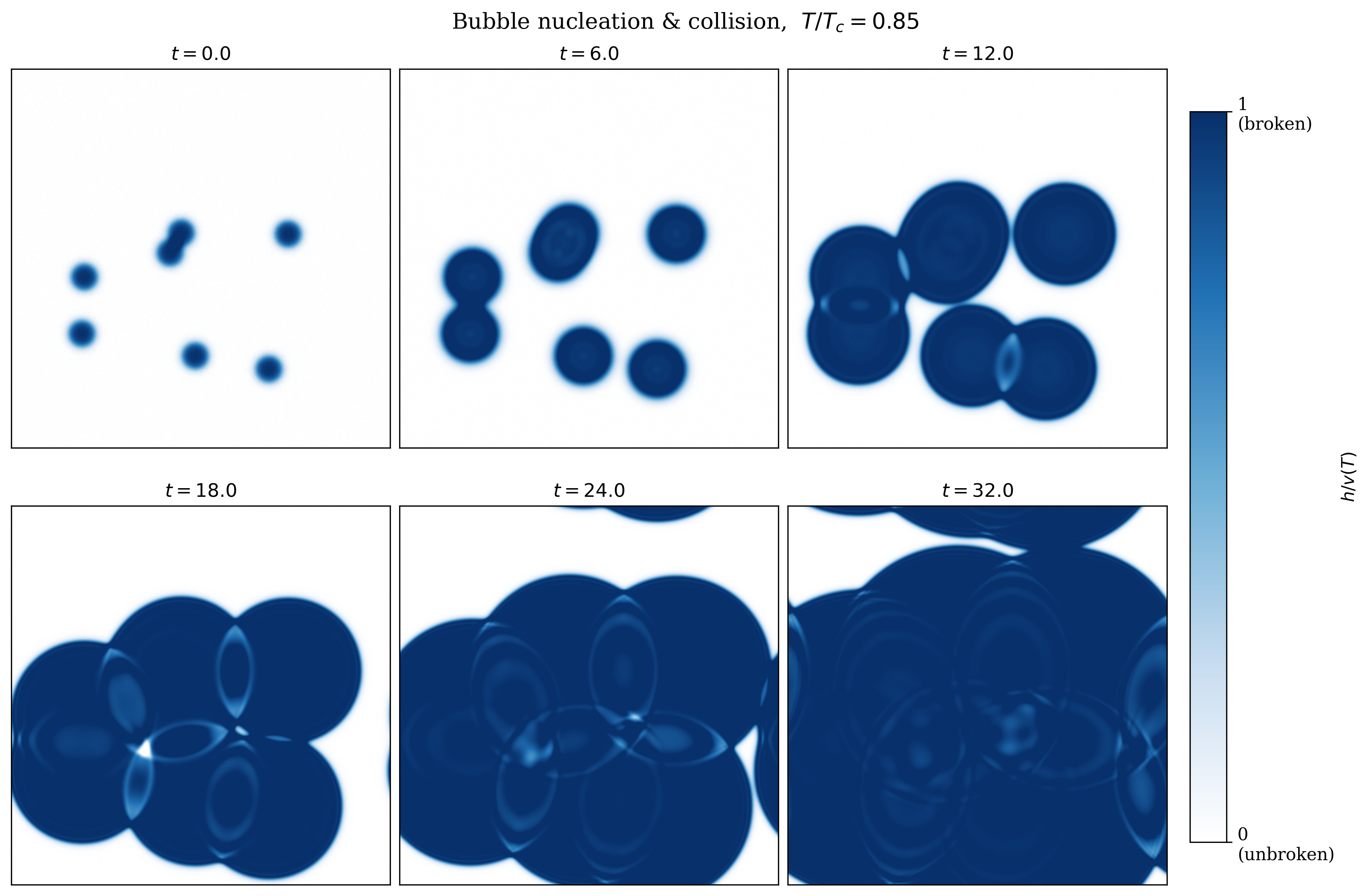}
    \caption{Schematic representation of the evolution, expansion and collision
of bubbles enclosing broken-symmetry regions. Notice that the
bubbles at t=0 start with an initial radius $R_c$, since at
subcritical radii the surface-energy contribution dominates,
causing them to shrink and disappear immediately.}
    \label{fig:bubble}
\end{figure}

First and foremost, the most important aspect of a first order
phase transition is the barrier which is created in effective
potential, between the true and the false vacuum regarding the
Higgs sector. The thermal barrier ensures that in the same
physical space coexist regions characterized by a broken symmetry,
i.e $h'\ne0$ and some defined by the unbroken symmetry in the
Higgs sector. Due to the fact that the true vacuum becomes more
energetically favored, since it minimizes the effective potential
$V'_{\text{eff}}(h', \phi', T') $, the bubbles containing region
with broken symmetry begin to expand rapidly while colliding with
one another vigorously, consequently leading to the creation of
stochastic gravitational wave background spectrum
\cite{Apreda:2001us,Schabinger:2005ei,Kusenko:2006rh,McDonald:1993ex,Chala:2018ari,Davoudiasl:2004be,Baldes:2016rqn,Noble:2007kk,Zhou:2020ojf,
Weir:2017wfa,Hindmarsh:2020hop,Han:2020ekm,Child:2012qg,Fairbairn:2013uta,LISACosmologyWorkingGroup:2022jok,Caprini:2015zlo,Huber:2015znp,
Delaunay:2007wb,Chung:2012vg,Barenboim:2012nh,Senaha:2020mop,Grojean:2006bp,Katz:2014bha,Alves:2018jsw,Athron:2023xlk}.
We shall use the framework of wall collision of the bubbles,
assuming the envelope approximation \cite{Kamionkowski:2015yta}
and the thin wall limit \cite{Ellis:2020awk}. This description is
schematically represented in Fig. \eqref{fig:bubble}.

This physical formalism allows us to compute the bubble nucleation
rate per unit of time and volume, $\Gamma(t)$, which is written
as,
\begin{equation}
    \Gamma(t)\thicksim T^4 \cdot  e^{\frac{-S_3}{T}}
\end{equation}
this term essentially expresses the free energy required for the
formation of a bubble. The term $S_3$ is defined as the three
dimensional Euclidean action derived from the fourth dimension
Euclidean action $S_E = \frac{1}{T} \int d^3x \left[
\frac{1}{2}(\nabla h)^2 + \Delta V_{\text{eff}}(h, T)
\right]$\footnote{The fourth dimension Euclidean action is derived
from the generic action of a scalar field $S_M =\int d^4x \left[
\frac{1}{2}(\partial^\mu h \partial_\mu h) - V_{\text{eff}}(h, T)
\right]$ by performing a Wick rotation t$\rightarrow$-i$\tau$.}.
Therefore $S_3=\int d^3x \left[ \frac{1}{2}(\nabla h)^2 + \Delta
V_{\text{eff}}(h, T) \right]$ \footnote{From now on the term
Euclidean action is used to refer to the three-dimensional
Euclidean action $S_3$.} and by considering a spherical spatial
symmetry, the Euclidean action is expressed as,
\begin{equation}
   S_3(T) = 4\pi \int_0^\infty dr \, r^2 \left[ \frac{1}{2} \left( \frac{dh}{dr} \right)^2 + \Delta V_{\text{eff}}(h_b, T) \right].
    \label{eq:S3_action}
\end{equation}
with an equation of motion,
\begin{equation}
  \frac{d^2 h_b}{dr^2} + \frac{2}{r}\frac{d h_b}{dr}
    = \frac{\partial\, \Delta V_{\text{eff}}(h_b, T)}{\partial h_b}
    \label{eq:S3_motion}
\end{equation}
The equation of motion regarding the action \eqref{eq:S3_action},
i.e. bounce equation \eqref{eq:S3_motion}, will be solved
analytically in this paper, since by taking advantage of the
thin-wall approximation, an estimation of the $S_3$ can be adopted
as follows \cite{Ellis:2020awk}:
\begin{equation}
    S_3=72\pi\frac{\sigma^3\xi_g^4}{a_*^2T^8}
    \label{s3_aprox}
\end{equation}
The parameters appearing in equation are connected to the
radiation energy density $\rho_r$ by the formula
$\rho_r=\frac{\pi^2}{30}g_*T^4$ \cite{Ellis:2020awk} and
$\xi_g=\sqrt{\frac{30 \ g_*}{\pi^2}}$, where $g_*$ refers to the
relativistic degrees of freedom at the percolation temperature.
The parameter $a_*$ appearing in equation \eqref{s3_aprox} defines
the strength of the phase transition, so it represents a different
method, besides calculating the $u_c/T_c$, to confirm whether a
first order phase transition is a strong one or a weak one.
Typical values of $a_*$ for describing the strength of the phase
transition are $a_*\thicksim O(0.01)$, which defines a weak phase
transition, $a_*\thicksim O(0.1)$ describe an intermediate phase
transition while values $a_*\thicksim O(1)$ or even larger
characterize a strong phase transition. The equation of $a_*$
regarding the parameters of the effective potential and energy
density is expressed as
\cite{Ellis:2020awk,Oikonomou:2024geq,Oikonomou:2026ugb}:
\begin{equation}
    \alpha_* =
    \left.
    \frac{\Delta V}{\rho_r}
    \right|_{T=T_*}
    \label{eq:alpha_stre}
\end{equation}
with $\Delta V$ denoting the effective potential difference
between the false and true vacuum states at the percolation
temperature $T_*$, whereas $\rho_r$ is the radiation density of
the relativistic plasma. The duration of the phase transition is
expressed in terms of the parameter $\frac{\beta}{H}$  where
$\beta^{-1} \thicksim \Delta t_{PT}$, in particular based on
\cite{Ellis:2020awk} the definition is formed as
$\frac{\beta}{H}=T'\frac{d}{dT'}(\frac{S_3}{T'})$ calculated at
$T_*$. Therefore,
\begin{equation}
    \frac{\beta}{H}=\frac{648\pi\cdot\sigma^3\cdot \xi_g^4}{a_*^2\cdot T^9}
    \label{eq:duration}
\end{equation}
Having defined the most important parameters for the gravitational
wave spectrum, namely $ \alpha_*$ and $\frac{\beta}{H}$, some more
details about the physical parameters of the bubbles will be
mentioned. First and foremost, the surface tension of the
generated bubbles is defined as,
\begin{equation}
    \sigma = \int_{h'_{false}}^{h'_{true}} dh' \sqrt{2V_{eff}^{SM}(h',0) - V_{true}}
    \label{eq:s_tension}
\end{equation}
which is calculated according to the mirror sector formalism of
the standard model developed in the former section. In addition,
another important parameter is,
\begin{equation}
   H_*R_*=(8\pi)^{1/3} v_w\frac{ H_*}{\beta}
\end{equation}
This essentially expresses the fractional size of the bubbles at
the time of their collision, relative to the size of the mirror
Hubble horizon at the epoch where bubble collision appears. The
critical radius at which bubbles are formed is given by
$R_c=\frac{2\sigma}{\Delta V}$. In this study the bubble wall's
radial velocity $v_w$ is set equal to 1, in order to comply with
NANOGrav's approach \cite{NANOGrav:2023gor}.

Proceeding now with the study of stochastic gravitational wave
background produced from bubble collision; in this section the
acknowledged formalism will be pointed out. First and foremost the
signal of gravitational waves is stochastic, meaning that it
originates from a collection of independent sources presenting
unpolarized profile while remaining isotropic. The amplitude of
gravitational waves is evaluated based on the energy density per
logarithmic frequency
$\Omega_{GW}(f)=\frac{1}{\rho_{tot}}\frac{d\rho_{GW}}{d\ lnf}$.
The energy density spectrum of stochastic gravitational wave
during the phase transition is given by
\cite{Apreda:2001us,Schabinger:2005ei,Kusenko:2006rh,McDonald:1993ex,Chala:2018ari,Davoudiasl:2004be,Baldes:2016rqn,Noble:2007kk,Zhou:2020ojf,
Weir:2017wfa,Hindmarsh:2020hop,Han:2020ekm,Child:2012qg,Fairbairn:2013uta,LISACosmologyWorkingGroup:2022jok,Caprini:2015zlo,Huber:2015znp,
Delaunay:2007wb,Chung:2012vg,Barenboim:2012nh,Senaha:2020mop,Grojean:2006bp,Katz:2014bha,Alves:2018jsw,Athron:2023xlk}:
\begin{equation}
    \Omega_b = \frac{\pi^2}{90} \frac{T_0^4}{M_p^2 H_0^2} g_* \left( \frac{g_{*,s}^{eq}}{g_{*,s}} \right)^{4/3} \tilde{\Omega}_b \left( \frac{\alpha_*}{1+\alpha_*} \right)^2 (H_* R_*)^2 \mathcal{S}(f/f_b)
    \label{eq:GW_spectrum}
\end{equation}
with $\tilde{\Omega}_b$=0.0049 and $\mathcal{S}(f/f_b)$ is defined as the spectral function of gravitational waves as:
\begin{equation}
    \mathcal{S}(x) = \frac{1}{\mathcal{N}} \frac{(a+b)^c}{(b x^{-a/c} + a x^{b/c})^c}
\end{equation}
The parameters $a,b$ influence the slope of the spectrum, whereas
parameter $c$ defines the width of the peak. $\mathcal{N}$ is the
normalization constant given by $\mathcal{N} = \left( \frac{b}{a}
\right)^{a/n} \left( \frac{nc}{b} \right)^c
\frac{\Gamma(a/n)\Gamma(b/n)}{n\Gamma(c)} $, where $n =
\frac{a+b}{c}$ and $\Gamma(z)$ represents the gamma function. In
this study the values $(a,b,c)=(1,1,3)$ are adopted accordingly to
the NANOGrav formalism \cite{NANOGrav:2023gor}. In Eq.
\eqref{eq:GW_spectrum} the parameter $\alpha_*$ once again refers
to the strength of the first order phase transition. The spectral
function is parameterized according to the peak frequency of the
gravitational wave spectrum, which is given by:

\begin{equation}
 f_b \simeq 48.5 \ nHz \ g_*^{1/2} \left( \frac{g_{*,s}^{eq}}{g_{*,s}} \right)^{1/3} \left( \frac{T_*}{\text{GeV}} \right) \frac{f_b^* R_*}{H_* R_*}
\end{equation}
with $f_b^*=0.58/R_*$ and $R_*$ being the mean bubble separation
at $T_*$, whereas $ g_* $ stands for the number of the
relativistic degrees of freedom contributing to the energy density
at $T_*$, $g_{*,s}$ and $g_{*,s}^{\mathrm{eq}}$ denote the entropy
contributing number of relativistic degrees of freedom at the
matter-radiation equality epoch. Finally it is important to
mention that  this paper follows the standard convention in which
the $\Omega_{GW}$ is multiplied by $h^2$ where $h=0.674 \pm
0.005$, and $H_0=100h$ kms$^{-1}$Mpc$^{-1}$.
\begin{figure}[htbp]
    \centering
\includegraphics[width=0.75\textwidth]{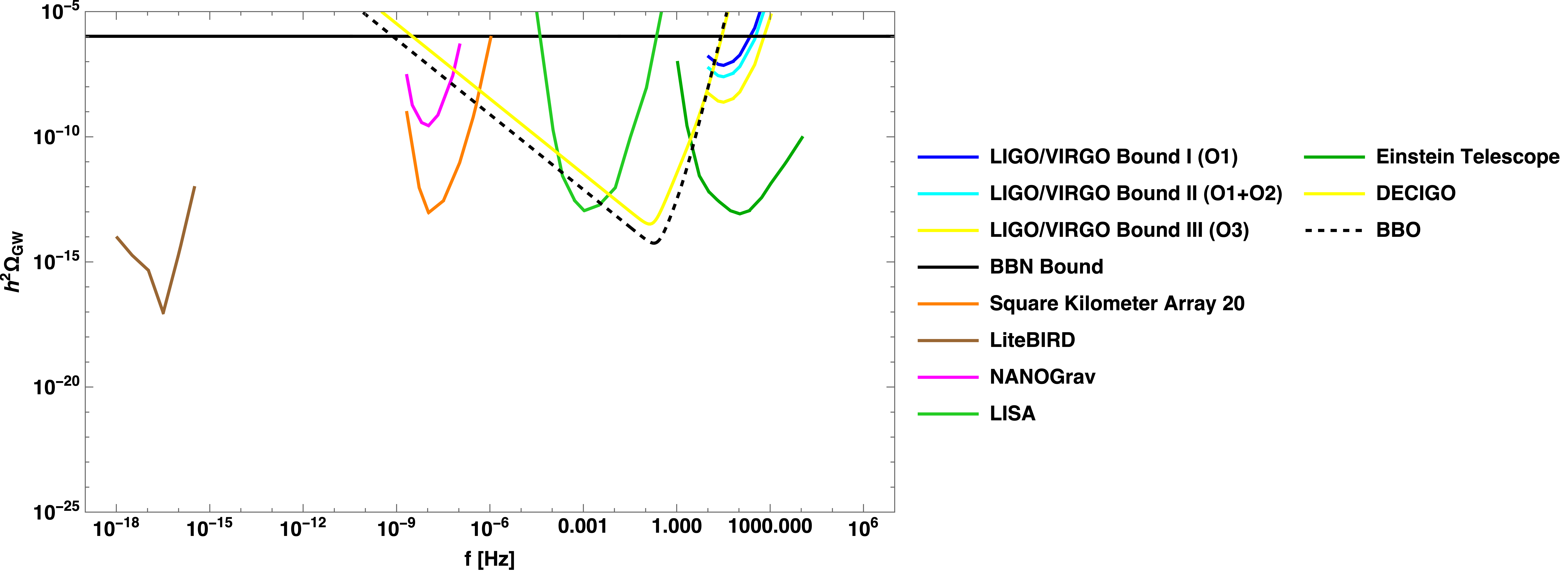}
\caption{Plot of the gravitational wave spectrum and sensitivity
bands for various detectors. In particular LISA interferometer
will exhibit a sensitivity around $10^{-4}$ to $10^{-1}$ Hz and
energy density spectrum will present a peak at
$\Omega_{GW}\thicksim 10^{-13}$. The  DECIGO will present a peak
at  $\Omega_{GW}\thicksim 2\cdot 10^{-15}$, around $0.1-10$ Hz.}
    \label{fig:signals}
\end{figure}


\subsection{Detection of Stochastic Gravitational Waves Generated from the mirror Electroweak Phase Transition}

Now that the general formalism for describing the gravitational
waves generated at the mirror electroweak phase transition at the
mirror sector is developed, we will proceed with the examination
of detectability of the three models mentioned in section III. The
energy density spectrum and the peak frequency of stochastic
gravitational waves are determined mainly by the difference of the
effective potential at the true and the false vacuum at the
percolation temperature, $\Delta V$,  which essentially
characterizes the strength of the phase transition $\alpha_*$
\eqref{eq:alpha_stre}. In addition the surface tension $\sigma$
\eqref{eq:s_tension} is a critical parameter which defines the
probability of forming bubbles, since it corresponds directly to
the energy needed for constructing a bubble. Having defined the
physical background of stochastic gravitational waves, in this
section we will address the gravitational waves produced from the
three models mentioned above. The sensitivity curves of the
various gravitational wave detectors are presented in Fig.
\ref{fig:signals}.

As discussed above, the sensitivity curves based on the models are
presented in Figs. \ref{fig:signal1}, \ref{fig:signal2} and
\ref{fig:signal3}.
\begin{itemize}
    \item The first model produces a detectable signal with a frequency peak
at $f_b \thicksim1.26\ mHz$ and an energy density spectrum
$h^2\Omega_{GW} \thicksim1.55 \cdot10^{-11}$. The signal is two
orders of magnitude larger than the minimum of the sensitivity
curve of the LISA and DECIGO/ BBO detectors, thus this signal may
be observable by the new generation detectors. The strength
parameter $\alpha_*\sim20$, therefore indicates a strong first
order phase transition. Finally the BBN bound is expressed as the
limit in the energy spectrum of the gravity waves at
$h^2\Omega_{GW} \thicksim1.1 \cdot10^{-6}$, thus the signal
remains valid and with BBN constraints.
    \begin{figure}[htbp]
        \centering
        \includegraphics[width=0.8\textwidth]{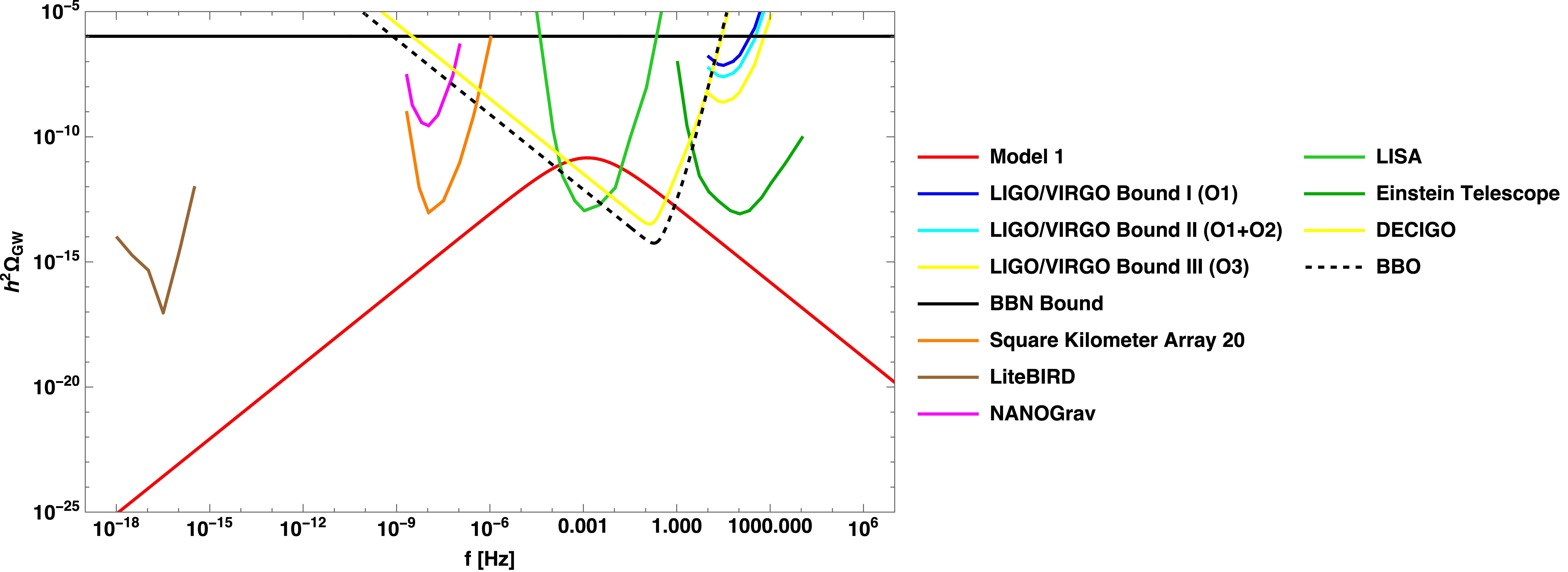}
        \caption{The predicted gravitational wave spectrum $h^2\Omega_{GW}(f)$ of
the model I (red solid line) which shows a peak at $f_b
\thicksim1.26\ mHz$ \&  $h^2\Omega_{GW} \thicksim1.55
\cdot10^{-11}$, versus the sensitivity curves of the present and
future detectors, namely: The LIGO/Virgo, the pulsar timing arrays
NANOGrav, the SKA (20 yr), LiteBIRD, LISA, DECIGO, and BBO and
Einstein Telescope. }
        \label{fig:signal1}
    \end{figure}
    \item The second model also lies within the detectability of
next-generation detectors, with a peak at $f_b \thicksim2.61\ mHz$
and energy density spectrum $h^2\Omega_{GW}
\thicksim2.2\cdot10^{-12}$. The strength parameter of the model is
calculated at $\alpha_*\sim22$ which indicates a strong phase
transition. This model is observable with LISA, DECIGO and BBO
detectors.
     \begin{figure}[htbp]
        \centering
        \includegraphics[width=0.8\textwidth]{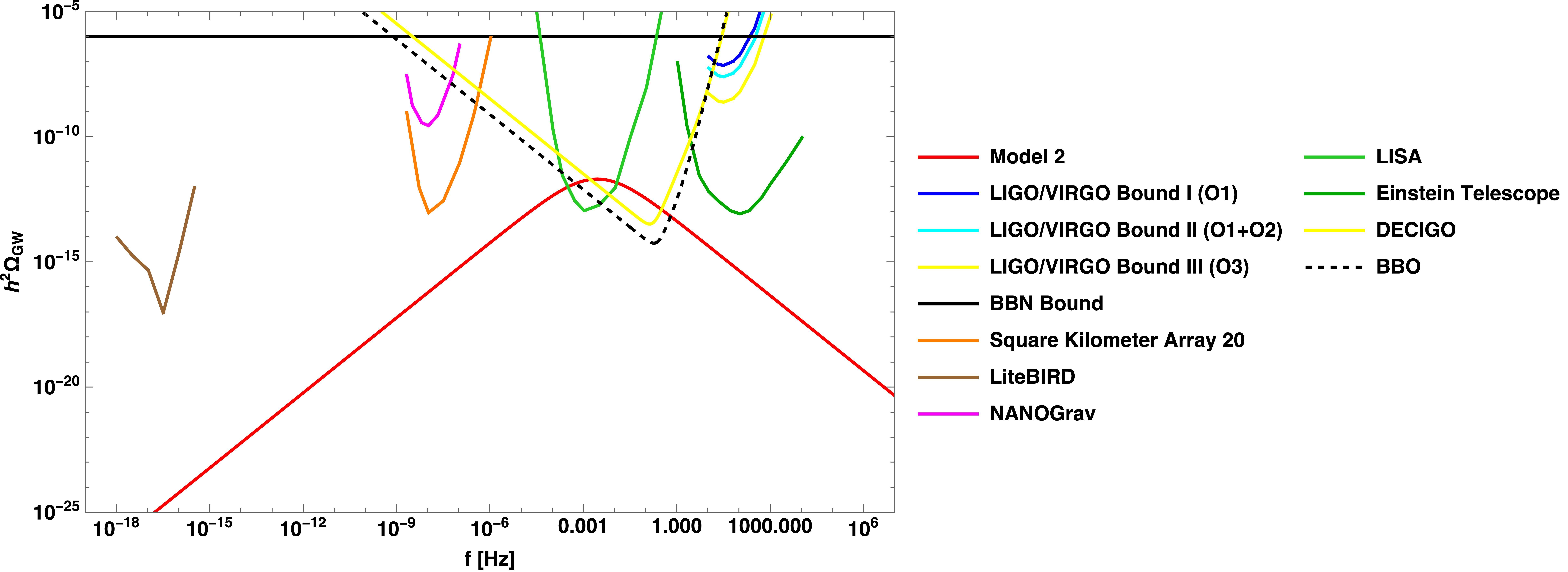}
        \caption{The predicted gravitational wave spectrum $h^2\Omega_{GW}(f)$ of
the model II, versus the sensitivity curves of the present and
future detectors, namely: The LIGO/Virgo, the pulsar timing arrays
NANOGrav, the SKA (20 yr), LiteBIRD, LISA, DECIGO, and BBO and
Einstein Telescope. For this model we have a spectral peak at $f_b
\sim 2.61\ \mathrm{mHz}$ and $h^2\Omega_{GW} \sim
2.2\cdot10^{-12}$. This signal remains within the sensitivity
curves of LISA, DECIGO, and BBO.}
        \label{fig:signal2}
    \end{figure}

    \item The third model presents a frequency maximum at  $f_b
\thicksim1.8\ mHz$ with a peak amplitude of $h^2\Omega_{GW}
\thicksim3.07\cdot10^{-12}$, and the strength parameter is
$\alpha_*\sim13.68$ which indicates that a strong first-order
phase transition generates the gravitational wave signals inside
the sensitivity curves of LISA, DECIGO, and BBO.
    \begin{figure}[htbp]
        \centering
        \includegraphics[width=0.8\textwidth]{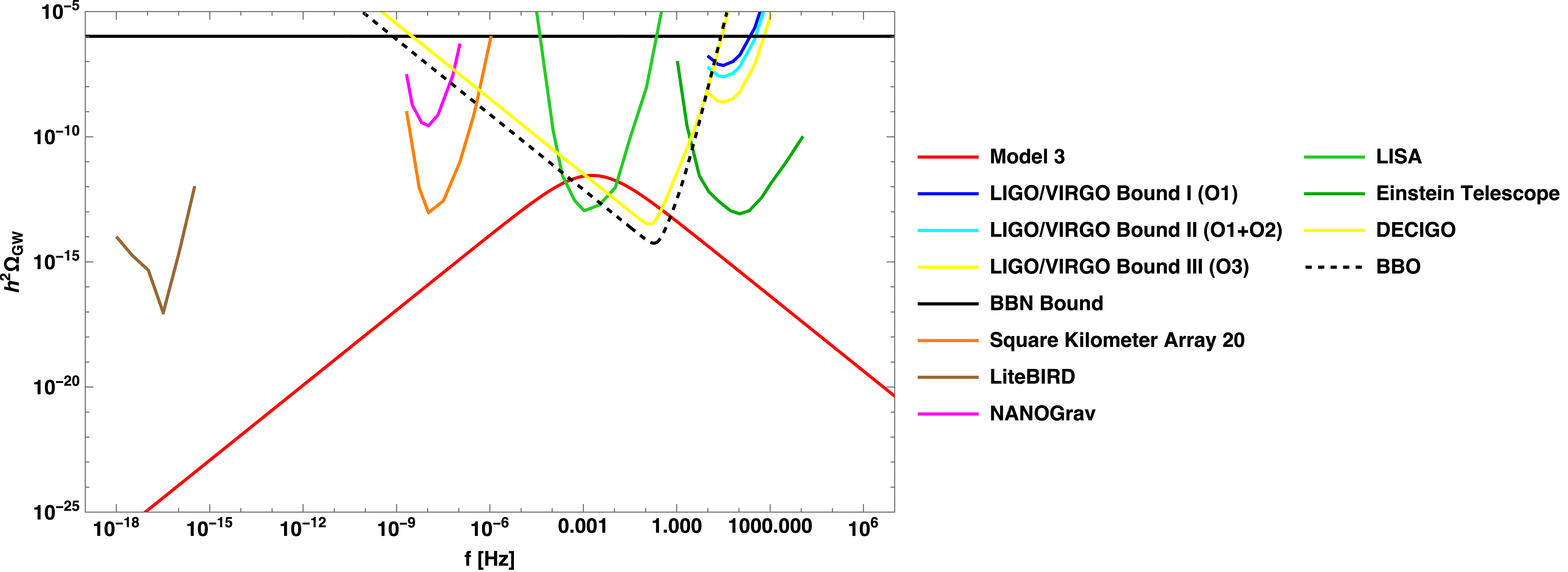}
        \caption{The predicted gravitational wave spectrum $h^2\Omega_{GW}(f)$ of
the model III, versus the sensitivity curves of the present and
future detectors, namely: The LIGO/Virgo, the pulsar timing arrays
NANOGrav, the SKA (20 yr), LiteBIRD, LISA, DECIGO, and BBO and
Einstein Telescope. In this model III the parameter $f_b
\thicksim1.8\ mHz$ \& $h^2\Omega_{GW} \thicksim3.07\cdot10^{-12}$
for this model. This signal is also detectable by the
next-generation interferometers.}
        \label{fig:signal3}
    \end{figure}

\end{itemize}
In summary, it has been shown in this analysis that all three high
scale mirror effective potential in finite temperature can produce
detectable stochastic gravitational wave signals, arising from the
mirror electroweak phase transition. The results of the three
models are presented in the Table \ref{tab:results}.

Now that the presentation of the three models is concluded, it is
important to note that we assumed that the main mechanism behind
the generation of gravitational waves is bubble-collision, in the
envelope approximation. This requires further justification, since
in the thermal first order phase transitions, the dominant
gravitational wave source is typically coming from the sound waves
generated in the surrounding plasma \cite{Athron:2023xlk}. The
criterion is focused on whether the bubble walls reach a terminal
velocity or run away, and at leading order, the friction exerted
by the plasma saturates at large wall velocities
\cite{Bodeker:2009qy}. So for strong transitions the released
vacuum energy cannot be transferred to the plasma, and instead it
accumulates in the walls of the bubbles. In terms of the parameter
$\alpha_\infty$, which is essentially the ratio of the released
energy over the radiation density \cite{Ellis:2020awk}, the three
models we presented satisfy $\alpha_* \sim 13\text{--}22 \gg
\alpha_\infty$, and the strong supercooling ($T'_* \simeq
0.3\,T'_c$) further dilutes the plasma.
\begin{table}[h!]
    \centering
    \renewcommand{\arraystretch}{1.8}
    \setlength{\tabcolsep}{12pt}
    \begin{tabular}{|c|c|c|c|c|c|}

        \hline
         & $u \ [GeV] $ & $ \displaystyle\frac{u_c'}{T_c'}$ & $\alpha_*$ & $f_{peak}\ [mHz]$ &$h^2\Omega_{GW}$ \\ \hline
        Model 1 & 6000  & 1.24 &  20     &    1.26   & $1.55 \cdot 10^{-11}$\\ \hline
         Model 2& 5000  & 1.02 &  22     &    2.61  & $2.2 \cdot 10^{-12}$\\ \hline
         Model 3& 4000  & 1.23 & 13.68   &    1.8   &  $3.07 \cdot 10^{-12}$\\ \hline
    \end{tabular}
    \caption{Overview of the results }
    \label{tab:results}
\end{table}


\section{The Stable Singlet as a Thermal Relic: An Accidental Dark Matter Candidate}
\label{sec:singletDM}

So far in this work we considered the mirror DM as one of the
components of the total DM in the Universe, and the mirror singlet
scalar had a simple role to make the dark phase transition
stronger. However that mirror singlet scalar has a more prominent
role than simply being a means to achieve a stronger first order
dark phase transition, since itself it can be a stable particle
and can be part of the total DM content in the Universe. There is
the important role that it can also be a link between the mirror
world and the real world, because it can be coupled to the mirror
Higgs and the real Higgs, an interesting scenario we did not
consider here though. We discuss this later on, and let us simply
focus on the case that the singlet scalar is only a part of the
mirror world.

The important feature of the mirror singlet scalar is that the
$\mathbb{Z}_2$ symmetry remains unbroken at all temperatures, thus
the mirror singlet $S'$ cannot decay,  however it can annihilate.
Therefore, the mirror singlet scalar can be stable and it can
survive as a thermal mirror relic of the mirror high scale
electroweak phase transition. Note that it is a real scalar, and
thus it does not have charge, save the $\mathbb{Z}_2$ parity,
therefore any annihilation of it will be strictly in pairs, and
not to particles or antiparticles. Being a stable particle, that
can annihilate, we can have a rough estimate of its relic
abundance, taking into account its annihilation cross section.
Since the singlet is coupled solely to the mirror Higgs, the
annihilation will involve the mirror Higgs particles, so we need
to check the scale of its mass. Note that we did not directly tune
the mass of the mirror singlet for purely DM purposes, but simply
we used it in order to achieve a stronger dark first order phase
transition. So by using the coupling value $\lambda_{HS}\simeq 6$
and the TeV scale $u'$, the singlet's mass is,
\begin{equation}
    m_S'^{\,2} = -\mu_S^2 + \lambda_{HS}\, u'^2
    \;\simeq\; \lambda_{HS}\, u'^2 \, ,
    \label{eq:mS_broken}
\end{equation}
where we used $\mu_S \ll \sqrt{\lambda_{HS}}\,u'$. So a rough
estimate for the Models I-III we developed in a previous section,
is $m_S' \simeq 9.6\text{--}15$TeV. The mirror singlet
annihilation process will involve the process $S'S' \to h'h'$,
which is controlled by the interaction
$\frac{\lambda_{HS}}{2}h'^2\phi'^2$ and the annihilation is an
$s$-wave process, with $\langle\sigma v\rangle$ being velocity
independent at leading order. Then, we expect roughly,
\begin{equation}
    \langle\sigma v\rangle \sim \frac{\lambda_{HS}^2}{32\pi\, m_S'^{\,2}}
    \sim 2\times 10^{-9}~\mathrm{GeV}^{-2}
    \simeq 2\times 10^{-26}~\mathrm{cm}^3\,\mathrm{s}^{-1}\, .
    \label{eq:sigmav}
\end{equation}
We can extend our analysis a bit further and seek for a rough
estimate of the relic abundance, so let us consider the freeze out
of the mirror singlet scalar in the mirror particles heat bath and
consequently the its relic abundance. Let the mirror singlet's
scalar number density be $n_S$, so the Boltzmann equation is,
\begin{equation}
    \frac{dn_S}{dt} + 3H n_S
    = -\langle\sigma v\rangle
    \left( n_S^2 - n_{S,\mathrm{eq}}^2 \right)\, .
    \label{eq:boltzmann}
\end{equation}
Note that the singlet scalar has no decays, only pair
annihilations to mirror Higgs particles, therefore the number
density evolution contains only quadratic dependence on $n_S$.
When the mirror temperature satisfies $T' \gg m_S'$, the mirror
singlet follows the equilibrium distribution and when the mirror
temperature $T'$ drops below $m_S'$, the equilibrium temperature
becomes Boltzmann-suppressed, $n_{S,\mathrm{eq}} \propto
e^{-m_S'/T'}$, hence the annihilation rate $\Gamma_{\rm ann} = n_S
\langle\sigma v\rangle$ eventually drops below the expansion rate.
The freeze out of the mirror singlet scalar occurs when $x_f' =
m_S'/T_f' \simeq 20-25$ and then, the comoving number density
becomes approximately constant. So we can integrate Eq.
\eqref{eq:boltzmann} from the instance $x_f'$ and beyond and then
we can obtain the relic abundance,
\begin{equation}
    \Omega_S h^2 \simeq 0.1
    \left( \frac{x_f'}{20} \right)
    \left( \frac{100}{g_*} \right)^{1/2}
    \left(
    \frac{3\times 10^{-26}~\mathrm{cm}^3\,\mathrm{s}^{-1}}
         {\langle\sigma v\rangle}
    \right)\, ,
    \label{eq:relic}
\end{equation}
and we used the reference value $\langle\sigma v\rangle_{\rm th}
\simeq 3\times 10^{-26}~\mathrm{cm}^3\,\mathrm{s}^{-1}$ which is
the canonical thermal cross section for a relic which is real
\cite{Jungman:1995df,Steigman:2012nb,Bertone:2004pz}. In Table
\ref{tab:relic} we gathered the rough estimations of our
calculations for the relic abundance of the mirror singlet, for
the models I-III we developed in the previous sections.
\begin{table}[htbp]
    \centering
    \renewcommand{\arraystretch}{1.3}
    \begin{tabular}{|l|c|c|c|}
        \hline
        & Model I & Model II & Model III \\ \hline
        $u'$ [TeV] & 6 & 5 & 4 \\ \hline
        $m_S'$ [TeV] & 15.0 & 12.4 & 9.6 \\ \hline
        $\langle\sigma v\rangle$ [$\mathrm{cm}^3\mathrm{s}^{-1}$] & $1.9\times 10^{-26}$  & $2.7\times 10^{-26}$ & $4.5\times 10^{-26}$ \\ \hline
        $\Omega_S h^2$ & 0.16 & 0.11 & 0.066 \\ \hline
    \end{tabular}
    \caption{Rough estimations of the relic abundance of the mirror singlet particle for the Models I-III of the previous section.}
    \label{tab:relic}
\end{table}
The rough estimate of Eq. (\ref{eq:relic}) indicates that although
in the Models I-III, the mirror DM abundance is small $\sim
0.012$, the singlet mirror scalar can account for the rest of the
DM. Hence, in this high scale mirror world which is extended by a
mirror singlet, all the DM of the Universe can be described by the
mirror world. This is quite interesting, but our result is merely
based on a tree order relic abundance result, without taking into
account radiative corrections due to the value of the
$\lambda_{HS}$ coupling. Hence for more concrete results a deeper
analysis is required, here we gave a rough estimation.

An interesting idea is to elevate the role of the singlet scalar
and instead of having it belonging to the mirror world only, it
can belong to both the mirror and the real world and have
couplings to both the high scale mirror Higgs and the ordinary
Higgs particle. Thus the mirror singlet can be an actual link
between the mirror and real world and in this scenario two phase
transitions could occur, one dark phase transition and one real
world phase transition. In this case, the frequencies that the
phase transitions may occur could be different, and thus this
pattern in the gravitational wave energy spectrum could be
distinctive. This scenario carries an additional constraint,
because if the singlet scalar belongs to the real world, then the
branching ration of the Higgs particle is constrained by the ATLAS
\cite{ATLAS:2020kdi} and CMS \cite{CMS:2018yfx} as \(BR_{inv} <
0.11 - 0.19\) at \(95 \% \) CL, and also
\cite{ParticleDataGroup:2022pth}, indicates that \(BR_{inv} <
0.107\) at \(95 \% \) CL \cite{ATLAS:2023tkt}.

The decay width of the Higgs to visible channels is \(\Gamma_{vis}
= 4.07\) MeV for a Higgs mass \(m_H = 125\) GeV. Therefore, if the
Higgs branching ratio to invisible particles is \(BR_{inv} <
0.19\), the invisible decay width of the Higgs boson upper bound
is,
\begin{equation}\label{decay_width_0.19}
    \Gamma (h \to \phi \phi) < 0.955 \text{ MeV}\, ,
\end{equation}
and the decay width is,
\begin{equation}\label{decay_width}
    \Gamma (h \to \phi \phi) = \frac{\lambda^2_{HS} \upsilon^2}{32 \pi m_H} \sqrt{1 - \frac{4
    m^2_S}{m^2_H}}\, .
\end{equation}
Therefore, the interacting coupling $\lambda_{HS}$ would be
constrained in the real and mirror world to be,
\begin{equation}\label{special_condition_coupling}
    \lambda_{HS} < \sqrt{ \frac{32 \pi m_H}{\upsilon^2}\left({1 - \frac{4 m^2_S}{m^2_H}} \right)^{-1/2}\Gamma_{m} (h \to \phi \phi)
    }\, .
\end{equation}
It would therefore be interesting to investigate this scenario,
and see both the real and mirror world effects on the primordial
gravitational waves energy spectrum. Of course, in this case, the
mirror world should not have such a high scale, as in the present
context. The scale of the mirror world $u'$ should be of the same
order as the one of the real world, or slightly larger or smaller.
We aim to address this issue in a forthcoming work.

\section{Conclusions}

In this work the focus was on studying a high scale mirror DM
world equipped with the symmetries of the SM, and also with a
singlet scalar field coupled to the mirror Higgs particle only. By
appropriately choosing the Yukawa couplings of the mirror DM
world, the mirror DM particles can also contains atoms and
elementary particles. The mirror DM world is completely detached
from the ordinary SM world and the two interact only
gravitationally. We showed that the high scale mirror SM equipped
with a singlet scalar extension, which has an unbroken
$\mathbb{Z}_2$ symmetry, produces detectable cosmological and
stochastic gravitational waves within the sensitivity of future
interferometers. Specifically, the values describing the
gravitational waves falls within the range of $f\thicksim O(1) \
mHz$ whereas the energy density is in the range of $h^2\Omega_{GW}
\thicksim O(2)\cdot10^{-12}$. These values can be directly
detected with LISA, DECIGO and BBO interferometers. Therefore in
the next few years such dark phase transition effects may be
observed by gravitational wave experiments. Ultimately, this work
establishes that bubble collisions generated during a strong
mirror world electroweak phase transition, thus a purely dark
phase transition, may serve as a primary candidate for producing a
stochastic gravitational wave background. An interesting
perspective we did not address in this work is the perspective of
having the singlet scalar belonging to both the ordinary SM world
and the mirror world simultaneously, and also that the mirror
scalar interacts simultaneously with both the ordinary and mirror
Higgs particle. If the scales of the ordinary and mirror world are
similar in order, this could potentially generate a distinctive
primordial gravitational waves pattern. We aim to address this
issue in a forthcoming article.

\section*{APPENDIX A: THERMAL SELF-ENERGY OF MIRROR PARTICLES }

In the finite temperature effective potential we developed in
section III, using the Arnold and Espinosa scheme, the term $
{V^\prime}^i_{ring} \left({m^\prime}^2_i(h', \phi'), T'\right)$ is
included into the formula of $ V'_{\text{eff}}(h', \phi', T')$, in
order to take into account only the zero Matsubara modes in the
resummation. The term $ {V^\prime}^i_{ring}
\left({m^\prime}^2_i(h', \phi'), T'\right)$ includes the
temperature-dependent self-energy  $ \Pi'_i(T')$ and this term is
different for each contributing field. The temperature dependent
self-energy for scalar fields is,
\begin{equation}
\Pi_h(T) = \Pi_\chi(T) = \left( \frac{3g'^2}{16} +
\frac{\tilde{g}'^2}{16} + \frac{y_t'^2}{4} + \frac{\lambda_H'}{2}
+ \frac{\lambda_{HS}}{12} \right) T'^2\, ,
\end{equation}
\begin{equation}
\Pi_S(T) = \left( \frac{\lambda_S}{4} + \frac{\lambda_{HS}}{3} +
\frac{\lambda v'^2}{2M^2} \right) T^2\, ,
\end{equation}
where $\chi$ corresponds to the mirror Goldstone bosons. Also the
self energy of longitudinal mirror gauge bosons is,
\begin{equation}
\Pi_{W_L'}(T') = \frac{11}{6} g'^2 T'^2
\end{equation}
and in the neutral sector, the thermal corrections mix the gauge
bosons $(A'^3_\mu,B'_\mu)$ with the basis components
$(A'^1_\mu,A'^2_\mu,A'^3_\mu,B'_\mu)$, hence the resummed masses
are obtained by rediagonalization of the corresponding $2\times2$
gauge boson mass matrix. The eigenvalues of this matrix are,
\begin{equation}
M_{Z_L}'^2 = \frac{1}{2} \left[ \frac{1}{4} \left(g^{\prime 2}
+ \tilde{g}^{\prime 2}\right) h^{\prime 2}
+ \frac{11}{6} \left(g^{\prime 2} + \tilde{g}^{\prime 2}\right) T^{\prime 2}
+ \sqrt{\left(g^{\prime 2} - \tilde{g}^{\prime 2}\right)^2
\left(\frac{1}{4} h^{\prime 2} + \frac{11}{6} T^{\prime 2}\right)^2
+ \frac{g^{\prime 2}\, \tilde{g}^{\prime 2}}{4}\, h^{\prime 4}} \, \right]
\end{equation}
\begin{equation}
M_{\gamma_L}'^2 = \frac{1}{2} \left[ \frac{1}{4} \left(g^{\prime
2} + \tilde{g}^{\prime 2}\right) h^{\prime 2} + \frac{11}{6}
\left(g^{\prime 2} + \tilde{g}^{\prime 2}\right) T^{\prime 2} -
\sqrt{\left(g^{\prime 2} - \tilde{g}^{\prime 2}\right)^2
\left(\frac{1}{4} h^{\prime 2} + \frac{11}{6} T^{\prime
2}\right)^2 + \frac{g^{\prime 2}\, \tilde{g}^{\prime 2}}{4}\,
h^{\prime 4}} \, \right]\, .
\end{equation}
When the zero-temperature limit is taken,  these eigenvalues
reduce to the tree-level masses of the mirror gauge bosons.

\end{document}